\documentclass[12pt]{article}
\usepackage{latexsym}
\usepackage{mathcomp}
\usepackage{graphicx}
\usepackage{graphics}
\usepackage{epstopdf}
\usepackage{epsfig}
\usepackage{graphicx}
\usepackage{float}
\usepackage{amsmath}
\usepackage{amsthm}
\usepackage{amssymb}
\usepackage{amscd}
\usepackage{amsfonts}
\usepackage{makeidx}
\usepackage{arcs}
\usepackage{mathrsfs}
\usepackage{bm}

 \newcommand{\beqn}{\begin{eqnarray}}
 \newcommand{\eeqn}{\end{eqnarray}}
 \newcommand{\be}{\begin{equation}}
 \newcommand{\ee}{\end{equation}}
 \newcommand{\ba}{\begin{array}}
 \newcommand{\ea}{\end{array}}

 \newcommand{\bfr}{\begin{flushright}}
 \newcommand{\efr}{\end{flushright}}
 \newcommand{\bfl}{\begin{flushleft}}

 \newcommand{\efl}{\end{flushleft}}

 \newcommand{\ds}{\displaystyle}

\newcommand{\no}{\noindent}
 \newcommand{\br}{|\kern-.25em|\kern-.25em|}
 \newcommand{\brr}{{|\kern-.15em|\kern-.15em|\kern-.15em}\,}
 \newcommand{\ddd}{\st{.\kern-.07em.\kern-.07em.}}

\newcommand{\bo}{{\hfill\loota}}
\newcommand{\loota}{\hbox{\enspace{\vrule height 7pt depth 0pt width
      7pt}}}

 \def\N{\mathbb{N}}                             

\def\R{\mathbb{R}}                              
 
 \def\Z{\mathbb{Z}}                                 

 \def\Re {{\rm Re\, }}                                       
 \def\Im {{\rm Im\,}}                                        

 \newcommand{\const}{\mathop{\rm const}\nolimits}
 \newcommand{\supp}{\mathop{\rm supp}\nolimits}

 \newtheorem{theorem}{Theorem}[section]

 \newtheorem{lemma}[theorem]{Lemma}
 \newtheorem{example}[theorem]{Example}
 \newtheorem{remark}[theorem]{Remark}

 \newtheorem{pro}[theorem]{Proposition}

\begin{document}

\title{Wave scattering by a periodic perturbation: embedded Rayleigh-Bloch modes and resonances}
\author{
\large{ P.  Zhevandrov$^1$},\\
\large{A.  Merzon$^2$},\\
\large{M.I. Romero Rodr\'iguez$^3$}\\
\large{and J.E. De la Paz M\'endez$^4$}\\
{\small{\it $^1$ Facultad de  Ciencias F\'\i sico-Matem\'aticas, Universidad Michoac{a}na}},\\[-2mm]
{\small  Morelia, Michoac\'{a}n, M\'{e}xico}\\[-2mm]
{\small{\it $^2$ Instituto de F\'\i sica y  Matem\'aticas, Universidad Michoac{a}na}},\\[-2mm]
{\small  Morelia, Michoac\'{a}n, M\'{e}xico}\\[-2mm]
{\small{\it $^3$ Facultad de Ciencias B\'asicas y Aplicadas, Universidad Militar Nueva Granada}},\\[-2mm]
{\small{Bogot\'a, Colombia}},\\[-2mm]
{\small{\it $^4$ Facultad de Matem\'aticas II, Universidad Aut\'onoma de Guerrero}},\\[-2mm]
{\small{Cd. Altamirano, Guerrero, M\'exico}}\\[-2mm]
{\small{\it E-mails}: pzhevand@gmail.com,}
{\small anatolimx@gmail.com,}\\[-2mm] {\small maria.romeror@unimilitar.edu.co,} {\small jeligio12@gmail.com}}
\date{}
\maketitle
\begin{center}
\date{\today}
\end{center}
\begin{abstract}
The scattering of quasiperiodic waves for a two-dimensional Helmholtz equation with a constant refractive index perturbed by a function which is periodic in one direction and of finite support in the other is considered. The scattering problem is uniquely solvable for almost all frequencies and formulas of Breit-Wigner and Fano type for the reflection and transmission coefficients are obtained in a neighborhood of the resonance (a pole of the reflection coefficient). We indicate also the values of the parameters involved which provide total transmission and reflection. For some exceptional frequencies and perturbations (when the imaginary part of the resonance vanishes) the scattering problem is not uniquely solvable and in the latter case there exist embedded Rayleigh-Bloch modes whose frequencies are explicitly calculated in terms of infinite convergent series in powers of the small parameter characterizing the magnitude of the perturbation.

\end{abstract}

\section{Introduction.}
\setcounter{equation}{0}

The appearance of trapped modes in unbounded domains
under perturbations has attracted a
lot of attention in both physical and mathematical literature in the recent past (see, e.g., the books \cite{Hurt, Lond, ExH} and references therein). In our previous paper \cite{arxiv} we studied this phenomenon for the Rayleigh-Bloch waves generated by a weak periodic perturbation. We considered a neighborhood of the cut-off of the continuous spectrum and constructed the discrete eigenvalue corresponding to a Rayleigh-Bloch wave trapped by the periodic structure; this eigenvalue lies outside the continuous spectrum. In the present paper we continue this investigation  for the frequencies embedded in the continuous spectrum close to the second cut-off. To specify the notation,
consider the Helmholtz equation
\begin{equation}\label{Helm}
-\nabla^2\Psi=\frac{\omega^2}{c^2(x,y)}\Psi,~~~\nabla=(\partial_{x},\partial_{y}),~~x,y\in\R,
\end{equation}
in the plane and assume that
\begin{equation}\label{cVAR}
c^{-2}(x,y)=1+\varepsilon f(x,y),\qquad 0<\varepsilon\ll1,
\end{equation}
where $f(x,y)$ is smooth, $T$-periodic with respect to $y$, $f(x,y+T)=f(x,y)$, and vanishes for $|x|>R$, see Fig.\;1.
 We will assume that the perturbation has  positive volume,
\begin{equation}\label{i7}
\int_0^T\int_{-\infty}^\infty f(x,y)\;dx dy>0.
\end{equation}

 \vspace{1cm}
\begin{figure}[htbp]
\centering
\includegraphics[scale=0.3]{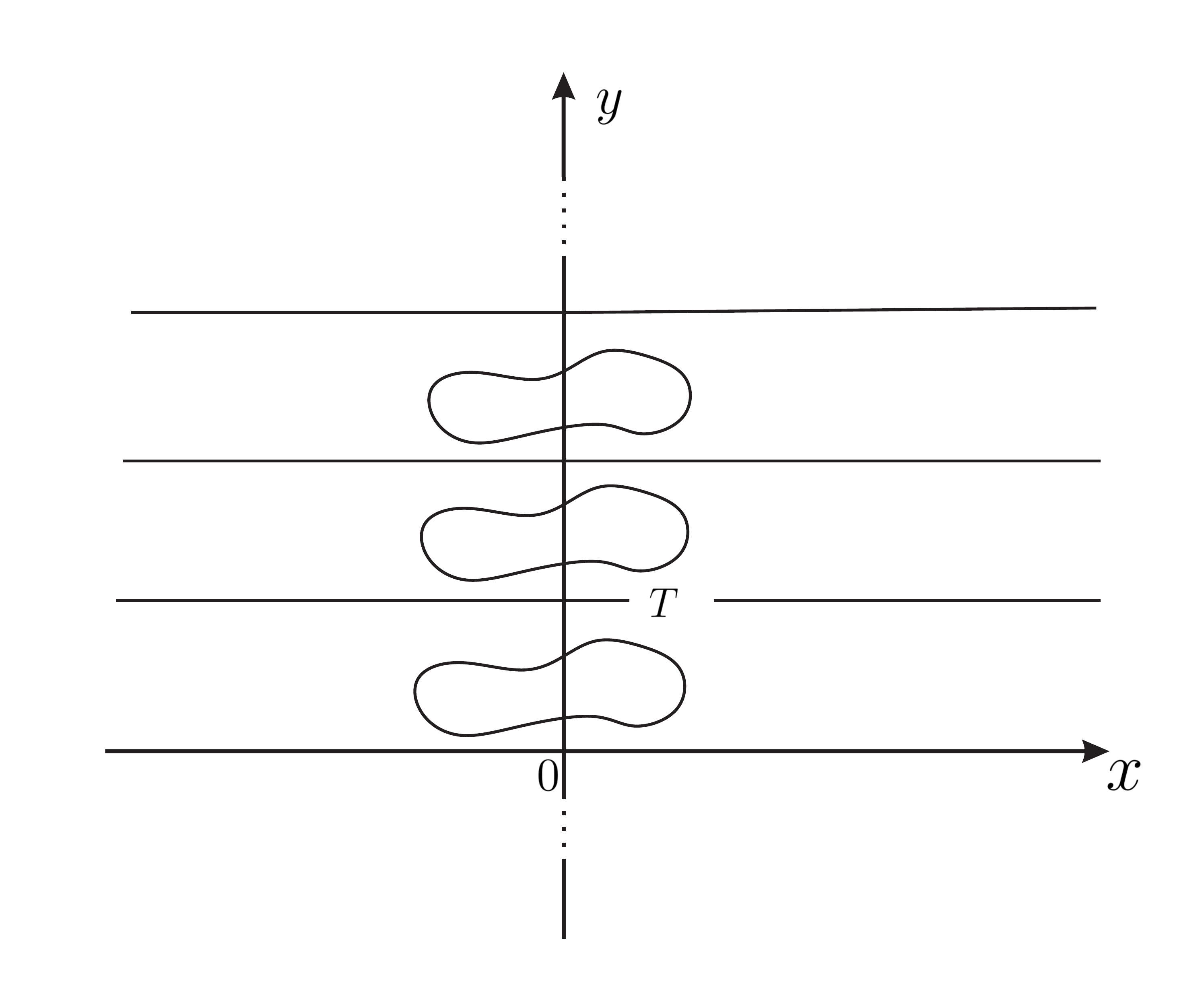}
\caption{The periodic structure}
\end{figure}

\vspace{.5cm}

We will be interested in quasiperiodic solutions of (\ref{Helm}) which have the form
\begin{equation*}
\Psi(x,y)=e^{i\beta y}\psi(x,y),
\end{equation*}
where $\beta$ is the ``wavenumber'', $\psi$ is $T$-periodic in $y$ and one can assume that $-\pi/T<\beta\leq \pi/T$ by the periodicity of $\psi$.
It is well-known that a periodic structure can support the so-called {\em Rayleigh-Bloch} (RB) {\em waves} which are solutions of the Helmholtz equation quasiperiodic in $y$ and decaying as $|x|\to \infty$.

\no There exists a vast amount of literature devoted to the study of RB waves in the setting (\ref{Helm})-(\ref{cVAR}) (without the assumption of the smallness of the perturbation, but for standing RB waves ($\beta=0$), see \cite{Ship1, Ship2, PtShV} and references therein), as well as for the setting of a half-plane with a periodic boundary (see \cite{W1, Bon, Naz} and references therein), and in the setting of a periodic array of solid obstacles (see \cite{Por Ev, Pet Mey, McI} and references therein).

\no Thus it seems of interest to investigate the existence of RB wave and associated phenomena (as, for example, the Fano resonance)
for a weak periodic perturbation of the refractive index (\ref{cVAR}) in the case of propagating modes ($\beta\neq0$). The construction of RB waves reduces to the solution of a boundary value problem in the strip
$$
\Pi=\R\times(0,T)
$$
for the Helmholtz equation with quasiperiodic boundary conditions; that is, we come to a waveguide problem. Obviously, without loss of generality, we can assume that $T=2\pi$. Problems of trapped modes in waveguides were quite extensively studied in numerous publications. We note the papers  \cite{Exner95, Bulla97}, where weakly perturbed quantum waveguides (i.e., perturbed Laplace operator with Dirichlet boundary conditions) were studied by means of the Birman-Schwinger technique (see \cite{ExH}), paper \cite{NazTMF} where embedded eigenvalues for this problem were considered, and the recent preprint \cite{ChN} where the Fano resonance was studied in connection with the problem of complete reflection/transmission (see also the bibliography in that paper).

\no The appearance of trapped modes in waveguides is frequently connected with the structure of the continuous spectrum of the unperturbed problem. In our case this spectrum is the ray $\omega_0^2\leq \omega^2$ (here $\omega_0^2=\beta^2$, see (\ref{coff})), which is divided by the numbers $\omega_1^2<\omega_2^2<\cdots <\cdots$ into segments on which the multiplicity is constant (for example, for $\beta\neq0,1/2$, on the interval $\omega_0^2<\omega^2<\omega_1^2=(1-|\beta|)^2$ this multiplicity is equal to $2$, that is, there exists one propagating mode; see Section\;2 below). As noted above, in our previous paper \cite{arxiv} we investigated the case when the first cut-off gives rise to a trapped mode under a perturbation; here we will be interested in trapped modes generated by the second cut-off. Since in the latter case there exists a propagating mode it is natural to pose the (standard) scattering problem. The reflection coefficient of the latter possesses a pole in the complex plane (resonance), and if the resonance is not real, the reflection and transmission coefficients ($\cal{R}$ and $\cal{T}$) undergo drastic changes in a neighborhood of the real part of the pole (the Breit-Wigner resonance, see \cite{Fried}; according to \cite{Naz}, these are analogues of the Wood anomalies). When the resonance becomes purely real, it can give rise to an eigenfunction, i.e., embedded Rayleigh-Bloch mode. Obviously, in this case the standard scattering problem does not possess a unique solution (an arbitrary multiple of the trapped mode, which decays at infinity, can be added to the solution without violating its existence). We note that it is possible to solve uniquely the scattering problem even in this case passing to the generalized formulation (see \cite{Ship1}), but we will not pursue this approach here since our principal goal consists in explicit formulas for the solution which is possible to obtain in the form of infinite series in powers of the small parameter $\varepsilon$.
It turns out that the resonance becomes purely real when the perturbation satisfies certain orthogonality condition which means that the length scale of the perturbation is connected with the wavelength of the propagating mode.
The principal difficulty of this problem  consists in the fact that the nonperturbed problem does not possess trapped modes, so that the standard regular perturbation theory is not applicable.

\no Our approach uses the main idea of the Birman-Schwinger method (i.e., the reduction of differential equations to integral ones) with simplifications and modifications (cf. \cite{ACZ}), \cite{MIR Zhev}).

\no In order to describe our results, we will need to introduce some notation. Denote by $f_{j}(x)$ the Fourier coefficients of the function $f(x,y)$ with respect to $y$:
\begin{equation}\label{i5}
f_j(x)=\int\limits_{0}^{2\pi} e^{-ij y} f(x,y)\;dy, \quad j\in\Z,
\end{equation}
and by $\tilde{f}_{j}(\xi)$ their Fourier transforms,
\begin{equation}\label{i6}
\tilde{f}_{j}(\xi)=\int e^{-i\xi x} f_{j}(x)\;dx;
\end{equation}
here and everywhere below integrals without limits mean the integration over the whole real axis.
We will assume that $0<\beta<1/2$ excluding the case $\beta=0$ (treated in \cite{Ship1, Ship2} in the case when the magnitude of the perturbation is  not small) and the case $\beta=1/2$.
These cases correspond to coalescence of the thresholds ($\omega_1$ and $\omega_2$ in the first case, and $\omega_0$ and $\omega_1$ in the second), and require a separate treatment.

\no Our main result consists in an explicit construction of a trapped mode in the case when the orthogonality condition mentioned above is satisfied for a certain value of $\beta$.
Our approach is based on the consideration of the scattering problem; as mentioned  above, the existence of trapped modes is connected with the nonuniqueness of solution to the latter. For simplicity, we will explain our results for positive $\beta$, $0<\beta<1/2$. We will also assume that $f(x,y)$ is even in both arguments,
\begin{equation}\label{sym}
f(-x,y)=f(x,y),\qquad f(x,-y)=f(x,y).
\end{equation}

Further, we will work in a neighborhood of the second threshold $\omega_1=1-\beta$ assuming that
\begin{equation}\label{i8}
\omega^2=\omega_1^2-\mu^2,\quad 0<\mu\ll 1.
\end{equation}
In this case (\ref{Helm}) (strictly speaking, the reduced problem in $\Pi$, see formulas (\ref{hh2}), (\ref{bc1}) below) possesses one propagating mode having the form
$$
e^{i\beta y\pm ik_0 x},\qquad k_0(\mu,\beta)=\sqrt{\omega^2-\beta^2}=\sqrt{1-2\beta-\mu^2}
$$
for $\varepsilon=0$. Denote the value of the wavenumber $k_0$ for $\mu=0$ as $\kappa$, $\kappa(\beta)=\sqrt{1-2\beta}$. The orthogonality condition has the form
\begin{equation}\label{i9}
\tilde{f}_1(\kappa)=0.
\end{equation}
Let $\beta_{00}$ solve (\ref{i9}), i.e., $\tilde{f}_1\Big(\kappa(\beta_{00})\Big)=0$.
We will prove in Section\;7 that if $\beta_{00}$ is such that  $\tilde{f}'_{1}\Big(\kappa(\beta_{00})\Big)\neq0$, then, for
\begin{equation}\label{mf}
\omega^2=(1-\beta)^2-\varepsilon^2\nu_{tr}^2,\qquad \nu_{tr}(\varepsilon)=\nu_{00}+O(\varepsilon), \qquad \nu_{00}=\frac{(1-\beta)^2}{4\pi} \tilde{f}_{0}(0),
\end{equation}
and for a certain $\beta=\beta_{tr}(\varepsilon)$ close to $\beta_{00}$, $\beta_{tr}(\varepsilon)=\beta_{00}+O(\varepsilon)$, there exists an RB wave. In Section\;7 we will define $\beta_{tr}, \nu_{tr}$ more precisely.

\no Apart from providing the information about the existence of RB waves, the consideration of the scattering problem sheds light on other interesting phenomena (e.g. the Fano resonance) connected with the existence of poles and zeros of the reflection and transmission coefficients studied, for example in \cite{Ship1}, \cite{Ship2}, \cite{ChN}. First of all, we note that in our setting the existence of trapped modes is due to the perturbation itself (that is, when $\varepsilon=0$ there are no trapped modes), while in \cite{Ship1}, \cite{ChN} the unperturbed problem possesses trapped modes. Of course, our results are different in what concerns the asymptotics of the reflection and transmission coefficients. Second, in contrast to \cite{Ship1}, we have  a three-parameter problem $(\varepsilon, \omega, \beta)$, while in \cite{Ship1} there are only two parameters $(\omega, \beta)$. Thus it may seem that our case is even more complicated. This is in part true, but the advantage of having a small parameter allows us to calculate explicitly all the quantities entering the formulas for the reflection and transmission coefficients in the form of expansions in powers of $\varepsilon$.

\no The consideration of the scattering problem allows us to obtain the information about the phenomena of total reflection and transmission. Namely, for $\beta$ such that $\tilde{f}_1(\kappa)\neq0$ (no trapped modes) and $\tilde{f}_0(2\kappa)\neq0$, there exist two values of $\nu=\mu/\varepsilon$ such that the reflection coefficient is zero (total transmission) at one of these values and the transmission coefficient is zero (total reflection) at the other (the Fano resonance). As the value of $\beta$ approaches the value for which there exists a trapped mode, these two points coalesce; on the $(\beta,\nu)$ plane they describe two curves that intersect tangentially at the point $(\beta_{tr}, \nu_{tr})$. The leading terms for the asymptotics of the reflection and transmission coefficients which are not regular in $\varepsilon$ (see(\ref{R,T})) admit a form resembling the famous Fano formula but, in our setting, these formulas differ from the Breit-Wigner formula by $O(\varepsilon)$ in the  additive sense. We note that the Breit-Wigner formula does not describe the existence of zeros mentioned above and has a symmetric form; the Fano formula, being a multiplicative asymptotics, describes these zeros and is asymmetric, the zeros lying  in the ``tails'' of the Breit-Wigner formula where its values are small. If the condition $\tilde{f}_0(2\kappa)\neq0$ is violated, we can guarantee that $\mathcal{R}$ does not have these additional zeros and even in the multiplicative sense has the Breit-Wigner form.

\no These results exemplify, for the case of small perturbations, the general results obtained in \cite{Ship2} in the sense that we calculate explicitly all the constants entering the corresponding asymptotics and indicate  the criteria for the existence of the Fano resonance. Note that in \cite{Ship1, Ship2, PtShV} only the existence of a trapped mode was proven for $\beta=0$ and the consideration of the propagating modes ($\beta\neq0$) is based on the assumption of their existence; we, in contrast, give an explicit construction of exact solutions describing them. Note also that we observe the Fano resonance also far away from the trapped mode (even when there are no trapped modes at all) while in \cite{Ship2} its existence is proven only for $\beta$ close to $\beta_{tr}$. Finally, apart from the two curves supporting the zeros of $\mathcal{R}$ and $\mathcal{T}$, we indicate the third curve on the $(\beta,\nu)$ plane which in the leading term in $\varepsilon$ coincides with the vertical line $\beta=\beta_{tr}$. Along this curve neither Fano nor Breit-Wigner resonances are present. In the last Subsection\;7.2 we obtain the Fano type formulas uniformly in the distance to the trapped mode $({\rm as}~ \beta\to \beta_{tr})$.

Let us gather together all the results mentioned above.

1) Let $\beta$ be such that $\tilde f_1(\kappa(\beta))\neq0$, $\kappa(\beta)=\sqrt{1-2\beta}$. Then the scattering problem has a unique solution and there are no embedded trapped modes. Further, let $\tilde f_0(2\kappa(\beta))\neq0$. Then $\cal{R}$ and $\cal{T}$ as functions of $\nu$, $\omega^2=\omega_1^2-\varepsilon^2\nu^2$, have zeros at the points $\nu=\nu_{a,b}$ (see (\ref{Na}), (\ref{Nb}) below; complete transmission and reflection, respectively) and the asymptotics of $\cal{R}$ and $\cal{T}$ are described by formulas of Breit-Wigner or Fano type (see, e.g., (\ref{SBW}) and  (\ref{F}) for $\cal{R}$). The width of the Breit-Wigner peak centered at the real part $\nu_{00}$ of the resonance (see (\ref{mf})) is of order of $\varepsilon$, while the distance of $\nu_a$ to the point $\nu_{00}$ is of order of 1. If  $\tilde f_0(2\kappa(\beta))=0$, then $\cal{T}$ still has a zero at the point $\nu_b$, while $\cal{R}$ does not have zeros in a neighborhood of order of $\varepsilon^{-1/2}$ of the point $\nu=\nu_{00}$; the asymptotics of $\cal{R}$ is described by the Breit-Wigner formula (\ref{SBW}). In this case there is no Fano anomaly since (\ref{SBW}) coincides with (\ref{F}). As far as we know, the criterion $\tilde f_0(2\kappa)\neq0$ for the Fano anomaly is new.

2) Let $\beta_{00}$ be such that $\tilde f_1(\kappa(\beta_{00}))=0$, $\tilde f'_1(\kappa(\beta_{00}))\neq0$. Then for the values $\beta_{tr},\nu_{tr}$ close to $\beta_{00},\nu_{00}$ there exists an embedded trapped mode whose dispersion relation is given by (\ref{mf}). Along a curve on the $(\beta,\nu)$ plane which in the first approximation coincides with the vertical line $\beta=\beta_{00}$, there are no Fano or Breit-Wigner anomalies. If the point $(\beta,\nu)$ is bounded away from this curve (but close to the point $\beta_{tr},\nu_{tr}$), then the reflection and transmission coefficients present the Fano or Breit-Wigner anomalies similarly to 1) above; the width of the Breit-Wigner peak is of order of $\varepsilon\Delta^2$, $\Delta\simeq\beta-\beta_{tr}(\varepsilon)$. In the case $\tilde f_0(2\kappa(\beta_{00}))=0$ the size of the neighborhood of the point $\nu=0$ free from zeros of $\cal{R}$ is of order of $\Delta$.


 The paper is organized as follow.
In Section\;2 we formulate the problem and describe the continuous spectrum of the unperturbed problem; in Section\;3 we formulate the scattering problem and in Section\;4 we solve it. In Section\;5 we obtain the Breit-Wigner type formulas for the reflection and transmission coefficients. In Section\;6 we investigate zeros of $\cal{R}$ and $\cal{T}$ (total transmission and reflection).  In Section\;7.1 we obtain the embedded trapped mode under the orthogonality condition (\ref{i9}) and discuss the scattering in this case, and in Section 7.2 we derive the asymptotics of $\cal{R}$ and $\cal{T}$ in a neighborhood of the point $(\nu_{tr},\beta_{tr})$.

\section{Statement of the problem.}
\setcounter{equation}{0}

Consider the Helmholtz equation
\begin{equation}\label{h1}
-\nabla^2\Psi(x,y)=\ds\frac{\omega^2}{c^2(x,y)}\Psi(x,y),~(x,y)\in\R^{2},~\omega\in\R,
\end{equation}
with a real smooth $2\pi$-periodic in $y$ refractive index,
\begin{equation*}
c(x,y+2\pi)=c(x,y),~(x,y)\in\R^{2}
\end{equation*}

Solutions of this equation that are quasiperiodic in $y$,
\begin{equation}\label{CP1}
\Psi(x,y+2\pi)= e^{2\pi i\beta}\Psi(x,y),\qquad -1/2<\beta\leq1/2,
\end{equation}
bounded in the whole plane and decaying as $|x|\to\infty$ are called {\it Rayleigh-Bloch waves} (RB waves) and exist only for certain values of the spectral parameter $\omega$.

We will be interested in the case of a weakly perturbed refractive index:
\begin{equation}\label{C1}
c^{-2}(x,y)= 1+\varepsilon f(x,y),~~0<\varepsilon\ll 1,
\end{equation}
where $f$ is a smooth $2\pi$-periodic in $y$ function vanishing for $|x|>R$ for some $R>0$
(see Fig.\;1, where the regions bounded by the closed curves mean the supports of $f$ when they are compact in the strip $0<y<2\pi$). We will restrict ourselves to the case of relatively low frequencies $\omega$ such that there exists only one propagating mode. For these frequencies one can pose a standard scattering problem (see (\ref{Psi1}) below).
It will turn out that the solution of this scattering problem exists and is unique under some additional conditions for $\beta$ and $f$ (see (\ref{mu0ne0})). In this case we obtain formulas of the Breit-Wigner type for the reflection and transmission coefficients, see Section\;4.  In the opposite case (roughly speaking, for certain values of $\beta$) the solution of the scattering problem is nonunique due to the existence of a trapped mode whose frequency is embedded in the  continuous spectrum.

\no Obviously, in order to construct the solution it is sufficient to find solutions of the following BVP in the strip
$\Pi:=\R\times(0,2\pi)$:
\begin{equation}\label{hh2}
-\nabla^2\Psi(x,y)=\omega^{2}\Big(1+\varepsilon f(x,y)\Big)\Psi(x,y),\qquad (x,y)\in\Pi,
\end{equation}
\begin{equation}\label{bc1}
\Psi(x,2\pi)=e^{2\pi i\beta}\Psi(x,0),~~~\Psi_{y}(x,2\pi)=e^{2\pi i\beta}\Psi_{y}(x,0).
\end{equation}

The continuous spectrum of problem (\ref{hh2}), (\ref{bc1}) is the ray
\begin{equation}\label{coff}
\omega^2\geq\omega_{0}^{2}=\beta^2
 \end{equation}
 and coincides with the continuous spectrum of the unperturbed problem ($\varepsilon=0$). The latter
 is divided by the cut-offs
\begin{equation}\label{CS}
\omega_{n,\pm}^2=(n\pm|\beta|)^2,~n=1,2,3,\dots,
\end{equation}
 into intervals where its multiplicity is constant; for $\beta\neq0,1/2$
 this multiplicity is equal to 2 on the first interval $\omega_0^2<\omega^2<\omega_1^2$, where
 \begin{equation}\label{om1}
 \omega_1^2=(1-|\beta|)^2,
 \end{equation}
 and is augmented by 2 when passing through the next threshold.
This follows from the explicit form of the plane waves
\begin{equation}\label{pwn}
\exp\{i(\beta+n)y+ikx\}
\end{equation}
satisfying (\ref{hh2})-(\ref{bc1}) for $\varepsilon=0$
with $\omega=\Omega_{n}(k)$ given by the dispersion relation
\begin{equation}\label{S1}
\Omega^2_{n}=k^2+\big(\beta+ n\big)^{2},\quad n\in\Z,\quad k\in\R.
\end{equation}
We will assume that $0<|\beta|<1/2$ excluding the cases $\beta=0,1/2$ since in the latter case the structure of the continuos spectrum is different (the jumps of the multiplicities are 4 instead of 2).

\begin{remark}\label{rm}
The function $\Psi$ in (\ref{hh2}), (\ref{bc1}) depends on four argument $x, y, \mu, \beta$. In the following we will sometimes omit the dependence of $\Psi$ on $\mu$ and $\beta$.
\end{remark}

For trapped modes, that is, for RB waves, the function $\Psi(x,y)$ in (\ref{hh2}), (\ref{bc1}) belongs to the space $H^{\beta}_{1} (\Pi)$ which is the completion of the space of smooth functions satisfying the boundary conditions from (\ref{bc1}) and vanishing for large $|x|$ with to respect to norm
\begin{equation*}
\parallel\Psi\parallel_{H_{1}^{\beta}(\Pi)}=\parallel\Psi\parallel_{L^{2}(\Pi)}+\parallel\nabla\Psi\parallel_{L^{2}(\Pi)}.
\end{equation*}
We understand  problem (\ref{hh2}), (\ref{bc1}) in the sense of the integral identity: for all smooth $\Phi$ satisfying (\ref{bc1}) and vanishing for $|x|\gg1$
\begin{equation}\label{it1}
\int_{\Pi} \nabla\Psi\cdot\nabla\overline{\Phi}\ dx dy=\omega^2\int_{\Pi}c^{-2}(x,y)\ \Psi\overline{\Phi}\ dxdy.
\end{equation}
This integral identity easily follows from (\ref{hh2}), (\ref{bc1}) after multiplying by $\overline{\Phi}$ and integrating over $\Pi$:
\begin{eqnarray*}
\begin{array}{ll}
\ds\int_{\Pi}\nabla^2\Psi\overline{\Phi}\ dxdy=\int_{\Pi} \nabla (\overline{\Phi}\ \nabla\Psi)\ dxdy\ -\int_{\Pi}\nabla \Psi\nabla\overline{\Phi}\ dxdy\\\\\hspace{2.6truecm}=\ds\int_{y=2\pi} \Psi_{y}\ \overline{\Phi}\ dx\ -\int_{y=0} \Psi_{y}\ \overline{\Phi}\ dx\ -\int_{\Pi} \nabla\Psi\cdot\nabla\overline{\Phi}\ dxdy.
\end{array}
\end{eqnarray*}
The first two terms in the last expression cancel out due to the boundary conditions in (\ref{bc1}). Moreover, the same argument shows that the operator corresponding to the problem (\ref{hh2}), (\ref{bc1}) is self-adjoint.
\begin{remark}\label{sa}
Strictly speaking, this calculation shows that the operator generated by the sesquilinear form in the left-hand side of (\ref{it1})  is self-adjoint in $H_{1}^{\beta}(\Pi)$ (see \cite{ExH}). Nevertheless the indication of this concrete space is not very important in what follows since we will construct a classical smooth solution of (\ref{hh2}), (\ref{bc1}).
\end{remark}

\section{Scattering problem. Reduction to an infinite system.}
\setcounter{equation}{0}

Let us assume that $\omega$ in (\ref{hh2}) satisfies the condition
\begin{equation}\label{w1}
\omega_0^2<\omega^2<\omega_1^2.
\end{equation}
This means that we are considering the frequencies between the first and the second cut-off, see (\ref{S1}). For these values of $\omega$ the unperturbed problem (\ref{hh2}), (\ref{bc1}) admits plane waves (\ref{pwn}) only for $n=0$ and with real $k$ satisfying (see (\ref{S1}))
$
\omega^2=k^2+\beta^2.
$
We will be interested (see e.g. \cite{Naz}) in the case when $\omega^2$ is close to the second cut-off $\omega_1^2$, that is,  we seek solutions of (\ref{hh2}), (\ref{bc1}) with
\begin{equation}\label{omega2}
\omega^2=\omega_1^2-\mu^2,~~~0<\mu\ll 1.
\end{equation}
Then for these $\omega$ there exist the plane waves (\ref{pwn}) with $k=\pm k_0$, $k_0=\sqrt{\omega^2-\beta^2}$,  and
hence we can pose the following scattering problem describing the scattering of a plane wave incident from the left. This problem consists in the construction of $\Psi(x,y)$ satisfying (\ref{hh2}), (\ref{bc1}) such that
\begin{equation}\label{Psi1}
  \Psi=\left\{
 \begin{array}{ll}
  e^{i\beta y+ik_0x}+\mathcal{R}\;e^{i\beta y-ik_0x}+o(1),& x\to -\infty,\\\\
 \mathcal{T} e^{i\beta y+ik_0x}+o(1),& x\to +\infty,
 \end{array}
 \right.
 \end{equation}
where  $\mathcal{R}$, $\mathcal{T}$ are the reflection and transmission coefficients.  More precisely (since $\Psi$ does not belong to $H_1^\beta$), formula (\ref{Psi1}) means that
$$
\Psi-\chi(x)\mathcal{T} e^{i\beta y+ik_0x}-\chi(-x)\Big(e^{i\beta y+ik_0x}+\mathcal{R} e^{i\beta y-ik_0x} \Big)\in H_1^{\beta}(\Pi),
$$
where $\chi(x)$ is a smooth cut-off function equal to $1$ for $x>1$ and to $0$ for $x<0$. The solution is still understood in the sense of the integral identity (\ref{it1}). We have $|\mathcal{R}|^2+|\mathcal{T}|^2=1$.
We will prove that the scattering problem admits a unique solution for almost all $\beta$ such that  $0<|\beta|<{1}/{2}$.

We seek $\Psi$ in the form
\begin{equation}\label{Psi}
\Psi(x,y,\mu)= e^{i\beta y}\sum_{n\in\Z}\Psi_{n}(x) e^{in y},~(x,y)\in\Pi,~\mu\in\R.
\end{equation}

\no Substituting (\ref{Psi}) in (\ref{hh2}) and carrying out the same procedure as was used in the derivation of equation (3.10) from \cite{arxiv}, we obtain
\begin{equation}\label{Eq4Psin}
-\Psi''_{m}+\Psi_{m}\Big[(\beta+m)^2-\omega^2\Big]=\varepsilon F_m,\quad m\in\Z,
\end{equation}
where $F_m=\ds\frac{\omega^2}{2\pi}\ds\sum_{n\in\Z}\Psi_{n} f_{m-n}$ (see (\ref{i5}) for the definition of $f_m$).
This is an infinite system of ordinary differential equations and in the following section we will obtain its solution.

\section{Solution of the infinite system. Conditions for its existence and uniqueness.}
\setcounter{equation}{0}
We will consider first the case of positive $\beta$, $0<\beta<1/2$, and below (see Subsection 4.3) we will explain how the case of negative $\beta$ should be treated.

\subsection{Solution for positive $\beta$.}

Let us solve system (\ref{Eq4Psin}). To this end we use the Green functions of equations (\ref{Eq4Psin}), considering separately the cases $m=0$, $m=-1$ and $m\neq 0, -1$.
\begin{enumerate}
\item $m=0$. We have
\begin{equation}\label{m0}
-\Psi''_{0}+\Psi_{0}\Big[-1+2\beta+\mu^2\Big]=\varepsilon F_{0}.
\end{equation}
The outgoing Green function $G_{0}$ for (\ref{m0}), has the form
\begin{equation}\label{G_00}
G_{0}(x,\mu)=\frac{i}{2k_{0}}\;e^{ik_{0}|x|},
\end{equation}
\begin{equation}\label{k_phi}
k_0=(1-2\beta-\mu^2)^{1/2}>0.
\end{equation}
\item $m=-1$. By (\ref{w1}), (\ref{Eq4Psin}), (\ref{om1}), we have
\begin{equation}\label{m-1}
-\Psi''_{-1}+\mu^2\Psi_{-1}=\varepsilon F_{-1}.
\end{equation}
The exponentially decaying Green function for (\ref{m-1}), is given by
\begin{equation}\label{G_1}
G_{-1}(x,\mu)=\frac{1}{2k_{-1}}\;e^{-k_{-1}|x|},~~k_{-1}=\mu.
\end{equation}
\item $m\neq0,-1$
\begin{equation}\label{m}
-\Psi''_{m}+\Psi_{m}\Big[(\beta+m)^2-\omega^{2}\Big]=\varepsilon F_{m}.
\end{equation}
Note that for $m\neq0,-1$ we have
\end{enumerate}
\begin{equation}\label{D1}
k_{m}:=\Big((\beta+m)^2-\omega_1^2+\mu^2\Big)^{\frac{1}{2}}>0
\end{equation}
and the exponentially decaying Green function for (\ref{m}) has the form
\begin{equation}\label{G_m}
G_{m}(x,\mu)=\frac{1}{2k_{m}}\;e^{-k_{m}|x|}, ~x\in\R.
\end{equation}
We see that the Green functions in the three cases considered above are quite different and have the following properties: in the first and third cases the Green functions are analytic in $\mu$ for small $\mu$, and in the second case the Green function is singular for $\mu\to0$. Moreover the quantities $k_{m}$ are real for real $\mu$ by (\ref{D1}).
\begin{remark}\label{Neg beta}
1. Note that for negative $\beta$, the singular (in $\mu$) Green function would correspond to $m=1$ instead of $m=-1$. We could repeat all the calculations below with this change, but prefer to circumvent this difficulty by means of passing to complex conjugates, see Subsection 4.3.

\no 2. Note also that at the values $\beta=0$ or $\beta=1/2$ {\it two} Green functions become singular in $\mu$ (for $m=\pm1$ in the first case and for $m=0,-1$ in the second); that is why we exclude these values of $\beta$.
\end{remark}
\no We seek the solution of (\ref{Eq4Psin}) in the form
\begin{equation}\label{Psi_0,Psi_m}
\Psi_0=e^{ik_{0} x}+G_{0}\ast A_{0},~~~\Psi_{m}=G_{m}\ast A_{m},~~m\neq0.
\end{equation}
Let us regularize $G_{-1}$. We have
\begin{equation}\label{G_1,1}
G_{-1}=G_{r}(x)+\frac{1}{2\mu},
\end{equation}
where
\begin{equation}\label{Gr}
G_{r}=G_{r}(x,\mu)=\frac{1}{2\mu}\Big(e^{-\mu|x|}-1\Big),~~x\in\R.
\end{equation}
Denote
\begin{equation}\label{Hn}
H_{n}=G_{n},~n\neq -1,~~~H_{n}=G_{r},~n=-1.
\end{equation}
\begin{remark}\label{rem}
The kernels $H_{n}$ are even by the definitions of $G_{m}$ (\ref{G_00}), (\ref{G_1}), (\ref{G_m}).
\end{remark}

\no In order to present the solution of system (\ref{Eq4Psin}) we still need to introduce several objects. Denote
\begin{equation}\label{gamma}
\gamma(\mu,\beta)=\frac{\omega^2}{4\pi}=\frac{1}{4\pi}\Big[(1-\beta)^2-\mu^2\Big].
\end{equation}
Sometimes we will omit the arguments $\mu, \beta$ for brevity.

\no Let $\mathcal{A}$ denote the space of vectors ${\bf A}=(\cdots, A_{-1}(x), A_{0}(x), A_{1}(x),\cdots)$ where $A_{j}(x)\in C[-R,R]$ with the norm
$$
\|{\bf A}\|_{\mathcal{A}}=\sum_{j\in\N}\Big(\sup_{x\in[-R,R]} |A_{j}(x)|\Big)^2.
$$
Obviously, $\mathcal{A}$ is a Banach space. Introduce the operator $\hat{T}:\mathcal{A}\to \mathcal{A}$ by the formula
\begin{equation}\label{hatT}
\Big(\hat{T} {\bf A}\Big)_{m}=2\gamma\sum_{n\in\Z}\Big(H_{n}\ast A_{n}\Big) f_{m-n}, \quad m\in\Z,
\end{equation}
where $({\bf A})_{m}$ means the $m$-th element of ${\bf A}$.
\begin{lemma}\label{lB}
{\bf i)}
The operator $\hat{T}:\mathcal{A}\to \mathcal{A}$ given by (\ref{hatT}) is bounded uniformly in $\mu$ for sufficiently small $\mu$ and
$$
\parallel\hat{T} {\bf A}\parallel_{\mathcal{A}}\leq \const \parallel{\bf A}\parallel_{\mathcal{A}}.
$$
{\bf ii)} If $A_{n}(x)$  decay rapidly as $n\to\infty$, i.e.,
$$
\sup_{x\in[-R,R]}|A_{n}(x)|=O\Big(|n|^{-N}\Big),~N\in\N,
$$
then the components of $\hat{T}{\bf A}$ also  decay rapidly as $n\to\infty$ uniformly in $\mu$ for sufficiently small $\mu$.
\end{lemma}
\no The proof of this fact is identical to the proof of Lemma\;3.5 from \cite{arxiv}.

\no Further, introduce the vectors ${\bf g}^{(1)}$ and ${\bf g}^{(2)}$ by
\begin{equation}\label{bfmgm1}
\Big({\bf g}^{(1)}\Big)_{m}= e^{ik_0x} f_{m},\quad \Big({\bf g}^{(2)}\Big)_{m}=f_{m+1},\quad m\in\Z.
\end{equation}
Consider the equation (dispersion relation for trapped modes, see Section\;7)
\begin{equation}\label{eqm4}
\mu-\varepsilon\gamma(\mu,\beta) F(\varepsilon, \mu,\beta)=0,
\end{equation}
where
\begin{equation}\label{eqm5}
F(\varepsilon, \mu, \beta)=\Bigg\langle \Bigg(\Big(1-\varepsilon\hat{T}\Big)^{-1} {\bf g}^{(2)}\Bigg)_{-1}\Bigg\rangle
\end{equation}
and the brackets $\langle\cdot\rangle$ mean the averaging, i.e.,
\begin{equation}\label{aver}
\Big\langle h \Big\rangle=\int\limits_{-\infty}^{\infty} h(x)\;dx.
\end{equation}
Obviously, $F$ is analytic in all its arguments for $\varepsilon$ and $\mu$ small and $0<\beta<1/2$.

\no By the Implicit Function Theorem, equation (\ref{eqm4}) possesses a unique root $\mu=\mu_0(\mu,\beta)$ which tends to zero as $\varepsilon\to0$. Denote
\begin{equation}\label{eqC}
C(\varepsilon, \mu,\beta)=\frac{2\varepsilon\mu\gamma(\mu,\beta)\;Q(\varepsilon, \mu,\beta)}{\mu-\varepsilon\gamma(\mu,\beta) F(\varepsilon, \mu,\beta)},
\end{equation}
where
\begin{equation}\label{Q}
Q(\varepsilon, \mu,\beta)=\Bigg\langle \Bigg(\Big(1-\varepsilon\hat{T}\Big)^{-1} {\bf g}^{(1)}\Bigg)_{-1}\Bigg\rangle.
\end{equation}

\no Denote
\begin{equation}\label{bfg_2}
{\bf g}={\bf g}^{(1)}+\frac{C}{2\mu}{\bf g}^{(2)}
\end{equation}
and let
\begin{equation}\label{eA}
{\bf A}=2\varepsilon\gamma\Big(1-\varepsilon\hat{T}\Big)^{-1} {\bf g}.
\end{equation}
\begin{theorem}\label{PR^pm}
Let $\mu-\varepsilon\gamma F\neq0$.
Then the function $\Psi$ defined by (\ref{Psi}) with $\Psi_n(x)$ defined by (\ref{Psi_0,Psi_m}) solves the scattering problem (\ref{hh2}), (\ref{bc1}), (\ref{Psi1}) with $\mathcal{R}$ and $\mathcal{T}$ given by
\begin{equation}\label{mathcalR}
\mathcal{R}=\ds\frac{i\varepsilon\gamma}{k_0}\;\frac{1}{\mu-\varepsilon\gamma F} \Bigg\lbrace\Big(\mu-\varepsilon\gamma F\Big) P^{+}+\varepsilon\gamma Q R^+\Bigg\rbrace,
\end{equation}
\begin{equation}\label{mathcalT}
\mathcal{T}=1+\frac{i\varepsilon\gamma}{k_0}\;\frac{1}{\mu-\varepsilon\gamma F}\Bigg\lbrace\Big(\mu-\varepsilon\gamma F\Big) P^{-}+\varepsilon\gamma Q R^{-}\Bigg\rbrace,
\end{equation}
where
\begin{equation}\label{PRpm}
  \left\{
 \begin{array}{ll}
  P^{\pm}(\varepsilon, \mu, \beta) &=\Bigg\langle \Bigg(\Big(1-\varepsilon\hat{T}\Big)^{-1} {\bf g}^{(1)}\Bigg)_{0} e^{\pm ik_0 x}\Bigg\rangle, \\\\
 R^{\pm}(\varepsilon, \mu, \beta)&=\Bigg\langle \Bigg(\Big(1-\varepsilon\hat{T}\Big)^{-1} {\bf g}^{(2)}\Bigg)_{0} e^{\pm ik_0 x}\Bigg\rangle.
 \end{array}
 \right.
 \end{equation}
\end{theorem}

\no {\bf Proof.} {\bf 1.} First of all, let us prove that $\Psi$ satisfies (\ref{hh2}), (\ref{bc1}). Substituting $\Psi$ in this system, we see that (\ref{bc1}) is automatically satisfied and in order that the Helmholtz equation be satisfied,  it is sufficient that $\Psi_m$ satisfy (\ref{Eq4Psin}). Substitute (\ref{Psi_0,Psi_m}), which defines $\Psi_m$, into (\ref{Eq4Psin}). Separating the summands corresponding to $n=0, -1$ in the right-hand side of (\ref{Eq4Psin}), we come to
\begin{equation}\label{vec A}
{\bf A}=2\gamma\varepsilon\Big({\bf g}^{(1)}+\frac{1}{2\mu}\langle A_{-1} \rangle g^{(1)}\Big)+\varepsilon\hat{T} {\bf A}
\end{equation}
by the definition of $\hat{T}$. Separating the $(-1)$st component in (\ref{vec A}) and averaging, we come to the following equation for $\langle A_{-1}\rangle$:
\begin{equation}\label{eqA_1}
\langle A_{-1}\rangle (\mu-\varepsilon\gamma F)=2\varepsilon\mu\gamma Q.
\end{equation}
Hence, $\langle A_{-1}\rangle$ coincides with $C$ in (\ref{eqC}). Thus (\ref{vec A}) is satisfied by the definition of ${\bf A}$ (see formula (\ref{eA})).

\no {\bf 2.} Let us derive the formulas for $\mathcal{R}$, $\mathcal{T}$. They follow directly from  (\ref{Psi}), (\ref{Psi_0,Psi_m}) and the fact that $A_{m}(\xi)$ have compact support and decay rapidly with respect to $m$ (see Lemma\;\ref{lB}) and since $G_{m}, m\neq0$, decay exponentially as $|x|\to\infty$ uniformly in $m$ by (\ref{G_1}), (\ref{G_m}).

\no Thus we see that the reflection  and transmission coefficients are given by the formulas
\begin{equation}\label{R,T}
\mathcal{R}=\frac{i}{2k_0}\int e^{ik_0 \xi} A_{0}(\xi)\;d\xi,~~~\mathcal{T}=1+\frac{i}{2k_0}\int e^{-ik_0 \xi} A_{0}(\xi)\;d\xi.
\end{equation}
By (\ref{eA}), we have
\begin{equation*}
\begin{array}{lll}
A_0(x) &=& 2\varepsilon\gamma\Bigg(\Big(1-\varepsilon\hat{T}\Big)^{-1} {\bf g}\Bigg)_{0}\\\\
&=& 2\varepsilon\gamma\Bigg(\Big(1-\varepsilon\hat{T}\Big)^{-1} {\bf g}^{(1)}\Bigg)_{0}+\ds\frac{2\varepsilon^2\gamma^2\; Q}{\mu-\varepsilon\gamma F}\cdot\Bigg(\Big(1-\varepsilon\hat{T}\Big)^{-1} {\bf g}^{(2)}\Bigg)_{0}.
\end{array}
\end{equation*}
Substituting in (\ref{R,T}), we obtain (\ref{mathcalR})-(\ref{PRpm}). ~~~$\bo$
\begin{remark}\label{ex}
 Theorem \ref{PR^pm} holds true when the expression $\mu-\varepsilon\gamma F$ does not vanish. This is the case if, for example,  $\rm{Im}\mu_0\neq0$, or if $\mu$ is bounded away from zero, $\mu\geq\rm{const}>0$ with $\const$ not depending on $\varepsilon$.
\end{remark}

\subsection{Investigation of the uniqueness condition, $\beta>0$.}

The condition
\begin{equation}\label{mu0ne0}
\Im\mu_0\neq0
\end{equation}
(see Remark \ref{ex}) is a condition for a solution of a nonlinear equation (\ref{eqm4}).
In this subsection we obtain simple sufficient conditions for the perturbation $f$ such that (\ref{mu0ne0}) holds.

\no Let us obtain these conditions.
\begin{lemma}\label{lImvar_0neq0}
The root $\mu_0(\varepsilon)$ of the equation (\ref{eqm4}) has the form
\begin{equation}\label{mu0-0}
\mu_{0}(\varepsilon,\beta)=a_1\varepsilon+a_2\varepsilon^2+\cdots,
\end{equation}
where
\begin{equation}\label{a1-1}
a_{1}=\gamma(0,\beta) \; \Big\langle f_{0}\Big\rangle,~~~a_{2}=\gamma(0,\beta)\; \Big\langle \Big(\hat{T}\big\vert_{\mu=0}\;{\bf g}^{(2)}\Big)_{-1}\Big\rangle.
\end{equation}
In particular,
\begin{equation}\label{Ia_1 a_2}
\Im a_1=0,~~\Im a_2=\frac{\gamma^2(0,\beta)}{2\kappa}\Big(|\tilde{f}_1(-\kappa)|^{2}+|\tilde{f}_1(\kappa)|^{2}\Big),
\end{equation}
where $\gamma$ is defined in (\ref{gamma}),
\begin{equation}\label{kappa}
\kappa=k_0\Big\vert_{\mu=0}=\sqrt{\omega_1^2-\beta^2}
\end{equation}
(see (\ref{G_00})).
\end{lemma}
\no The proof of this lemma can be found in Appendix\;1.

\begin{remark}\label{kap,-kap}
We see that the condition $\Im\mu_0(\varepsilon)\neq0$ is satisfied if the Fourier transform of the first Fourier coefficient of the perturbation does not vanish at the point $\kappa$ or $-\kappa$ which are the wave numbers  (up to $O(\mu)$) of the propagating mode.
\end{remark}
\medskip

\no Lemma \ref{lImvar_0neq0} implies that under the condition $\tilde{f}_1(\kappa)\neq0$ or $\tilde{f}_1(-\kappa)\neq0$ the parenthesis in the left-hand side of (\ref{eqA_1}) does not vanish by (\ref{mu0ne0}), (\ref{Ia_1 a_2}) and (\ref{mu0-0}) and the definition of $\mu_0(\varepsilon,\beta)$. Hence we obtain the vector $\bf{A}$ by means of (\ref{eA})
where $\bf{g}$ is given by (\ref{bfg_2}) and the constant $\langle A_{-1}\rangle$ is given by (\ref{eqA_1}). Thus, we obtain also the unique solution $\Psi$ of the scattering problem (\ref{Psi1}), given by the formulas (\ref{Psi_0,Psi_m}), (\ref{eA}).

\subsection{Solution for negative $\beta$.}

For negative $\beta$, $-1/2<\beta<0$, $\hat{\beta}:=-\beta>0$ we consider the solution $\Phi$ of (\ref{hh2}) satisfying the conditions
\begin{equation}\label{per1}
\Phi(x,2\pi)=e^{2\pi i\hat{\beta}}\Phi(x,0),\quad \Phi_{y}(x,2\pi)=e^{2\pi i\hat{\beta}}\Phi_{y}(x,0)
\end{equation}
and satisfying
\begin{equation}\label{seat1}
  \Phi=\left\{
 \begin{array}{ll}
  e^{+i\beta y-ik_0x}+\overline{\mathcal{R}}\;e^{+i\beta y+ik_0x}+o(1),& x\to -\infty,\\\\
 \overline{\mathcal{T}} e^{+i\beta y-ik_0x}+o(1),& x\to +\infty,
 \end{array}
 \right.
 \end{equation}
 and $ \overline{\mathcal{R}}$, $ \overline{\mathcal{T}}$ are to be determined. Then the function $\Psi=\overline{\Phi}$, with $\mathcal{R}$,
 $\mathcal{T}$ being the complex conjugates to $\overline{\mathcal{R}}$, $\overline{\mathcal{T}}$ will satisfy (\ref{hh2}), (\ref{bc1}) and (\ref{Psi1}). Thus, it is sufficient to construct the solution of (\ref{hh2}), (\ref{per1}), (\ref{seat1}).

 This is done exactly in the same way as above, the only difference consisting in the choice of the Green function $G_0$, which should be given by the same formula (\ref{G_00}) where instead of the positive root $k_0=(1-2\hat{\beta}-\mu^2)^{1/2}$ one has to take $k_0=-(1-2\hat{\beta}-\mu^2)^{1/2}$. Also, in all the formulas $\beta$ should be changed to $\hat{\beta}$ and $k_0$ should be changed to $-k_0$. The statement of Theorem\;\ref{PR^pm} holds true for $\Psi=\overline{\Phi}$, where $\Phi$ is constructed as indicated above.

\section{Asymptotics of $\mathcal{R}$ and $\mathcal{T}$.}
\setcounter{equation}{0}

For simplicity, we will consider only positive $\beta\in(0,1/2)$ (negative $\beta$ can be treated as in Subsection 4.3). Moreover, we will be working in a small $\varepsilon$-neighborhood of the second  threshold $\omega_1$ and thus it is convenient to introduce a new variable $\nu$ as $\mu=\varepsilon\nu$, $0<\nu<\const$. Our goal is to obtain asymptotic formulas for $\mathcal{R}$ and $\mathcal{T}$ as $\varepsilon\to 0$ when the imaginary part of the root $\mu_0(\varepsilon,\beta)$ of equation (\ref{eqm4}) does not vanish, or, what is the same, the imaginary part of the root $\nu_0(\varepsilon,\beta)$ of the equation
\begin{equation}\label{eqnu}
\nu-\gamma(\varepsilon\nu,\beta)F(\varepsilon,\varepsilon\nu,\beta)=0
\end{equation}
does not vanish. We will need the asymptotics for all the functions entering (\ref{mathcalR}), (\ref{mathcalT}). It is straightforward to obtain, using (\ref{eqm5}), (\ref{Q}), (\ref{PRpm}), that
\begin{equation}\label{PQR}
    \left.\ba{rcl}
            F(\varepsilon,\varepsilon\nu,\beta) &=& \tilde{f}_{0}(0)+O(\varepsilon+\varepsilon|\nu|), \\\\
 Q(\varepsilon,\varepsilon\nu,\beta)&=& \tilde{f}_{-1}(-\kappa)+O(\varepsilon+\varepsilon|\nu|)=\overline{\tilde{f}_1}(\kappa)+O(\varepsilon+\varepsilon|\nu|)\\\\
P^{+}(\varepsilon,\varepsilon\nu,\beta)&=& \tilde{f}_{0}(-2\kappa)+O(\varepsilon+\varepsilon|\nu|)=\overline{\tilde{f}_0}(2\kappa)+O(\varepsilon+\varepsilon|\nu|)\\\\
P^{-}(\varepsilon,\varepsilon\nu,\beta)&=& \tilde{f}_{0}(0)+O(\varepsilon+\varepsilon|\nu|)\\\\
R^{\pm}(\varepsilon,\varepsilon\nu,\beta)&=& \tilde{f}_1(\mp\kappa)+O(\varepsilon+\varepsilon|\nu|).
        \ea
\right|
\end{equation}
We will investigate the behavior of $\mathcal{R}$  and $\mathcal{T}$ as functions of $\nu$ in a neighborhood of the point
$$
r_0=\Re\nu_0(\varepsilon,\beta)=a_1+\varepsilon\Re a_2+\cdots,
$$
setting $\nu=r_0+\delta$ with $\delta$ so small that $\nu>0$ (see (ref{mu0-0}), $r_0=\Re\mu_0/\varepsilon$).

\no Denote
\begin{equation*}
d_{\pm}:=\tilde{f}_{1}(\pm\kappa),\quad \quad \Gamma:=\frac{\gamma^2(0,\beta)}{\kappa}\Big(|d_{+}|^{2}+|d_{-}|^{2}\Big).
\end{equation*}
\begin{theorem}\label{teo}
Let $\Gamma\neq0$ and
\begin{equation}\label{delta}\delta\in\Big(-\delta_0(\varepsilon), \delta_0(\varepsilon)\Big), \qquad \varepsilon\delta_0^2(\varepsilon)\xrightarrow[\varepsilon\to 0]{}0.\end{equation}
Then
\begin{equation}\label{R}
            \mathcal{R}(\varepsilon,\delta)=W\Big(\frac{\delta}{\varepsilon}\Big)\Bigg(\frac{\gamma^2(0,\beta)}{\kappa}\overline{d}_{+} d_{-}+\delta\overline{\tilde{f}_0}(2\kappa)\frac{\gamma(0,\beta)}{\kappa}+ O(\varepsilon+\varepsilon\delta^2)\Bigg),
\end{equation}
\begin{equation}\label{T}
\mathcal{T}(\varepsilon,\delta)=W\Big(\frac{\delta}{\varepsilon}\Big)\Bigg(\frac{\delta}{i\varepsilon}+
\frac{\gamma^2(0,\beta)}{2\kappa}\Big(|d_+|^2-|d_{-}|^2\Big)+ \delta\tilde f_0(0)\frac{\gamma(0,\beta)}\kappa+O(\varepsilon+\varepsilon\delta^2)\Bigg),
\end{equation}
where
$$
W(z)=\frac{i}{z-i\Gamma/2}.
$$
\end{theorem}
\no {\bf Proof.} Consider the expression
\begin{equation}\label{d}
\mu-\varepsilon\gamma(\mu,\beta) F(\varepsilon,\mu,\beta)=\varepsilon\Big(\nu-\gamma(\varepsilon\nu,\beta) F(\varepsilon,\varepsilon\nu,\beta)\Big)
\end{equation}
entering (\ref{mathcalR}), (\ref{mathcalT}). We have by the Lagrange formula
\begin{equation}\label{Lag}
\nu-\gamma(\varepsilon\nu,\beta) F(\varepsilon,\varepsilon\nu,\beta)=\Big(\nu-\nu_0(\varepsilon,\beta)\Big) \Big(1+O(\varepsilon)\Big).
\end{equation}
The proof of this formula can be found in Appendix\;2.
Hence the asymptotics of the denominator in (\ref{mathcalR}), (\ref{mathcalT}) has the form
\begin{equation}\label{nunu emp}
\nu-\gamma F=(\delta-i\varepsilon\Gamma/2+O(\varepsilon^2))(1+O(\varepsilon)).
\end{equation}
 Let us  obtain an asymptotics of the numerators in these formulas. We have
$$
\big(\mu-\varepsilon\gamma F\big)P^{+}+\varepsilon\gamma Q R^{+}=\varepsilon\Big(\delta\overline{\tilde{f}_0}(2\kappa)+\gamma(0,\beta) \overline{\tilde{f}_1}(\kappa)\tilde{f}_1(-\kappa)+O(\varepsilon+\varepsilon\delta^2)\Big)
$$
since
$$
P^{+}\Big(\varepsilon,\varepsilon(\delta+\Re\nu_0),\beta\Big)=\overline{\tilde{f}_0}(2\kappa)+O(\varepsilon+\varepsilon|\delta|),\quad
Q\Big(\varepsilon,\varepsilon(\delta+\Re\nu_0),\beta\Big)=\overline{\tilde{f}_1}(\kappa)+O(\varepsilon+\varepsilon|\delta|),\quad
$$
$$R^{+}\Big(\varepsilon,\varepsilon(\delta+\Re\nu_0),\beta\Big)={\tilde{f}_1}(-\kappa)+O(\varepsilon+\varepsilon|\delta|),
$$
by (\ref{PQR}). Thus for $\mathcal{R}$ we obtain
$$
\mathcal{R}=i\frac{\varepsilon\Big(\delta \overline{\tilde{f}_0}(2\kappa)\gamma(0,\beta)/\kappa+\gamma^2(0,\beta)\overline{d}_{+} d_{-}/\kappa+O(\varepsilon+\varepsilon\delta^2)\Big)}{\delta-i\varepsilon\Gamma/2+O(\varepsilon^2)}\Big(1+O(\varepsilon)\Big)
$$
It is easy to see that, since $\Gamma\neq0$,
\begin{equation}\label{den}
\frac{1}{\delta-i\varepsilon\Gamma/2+O(\varepsilon^2)}=\frac{1}{\delta-i\varepsilon\Gamma/2}\Bigg(1+O(\varepsilon) \Bigg),
\end{equation}
and hence we obtain (\ref{R}) as claimed. Similarly, we obtain (\ref{T}).  ~$\bo$

\section{Zeros of the reflection and transmission coefficients and the Fano resonance.}
\setcounter{equation}{0}
From now on we will assume the symmetry conditions (\ref{sym}) from the Introduction and will work with the variable $\nu=\mu/\varepsilon$. In this case we have that all the functions $f_{j}(x)$ are real and even and $f_{-j}(x)=f_{j}(x)$. Moreover, the vector
\begin{equation}\label{Y}
{\bf Y}=(1-\varepsilon\hat{T})^{-1} {\bf g}^{(2)}
\end{equation}
is also even in $x$ since $\hat{T}$ preserves the evenness (since all the kernels $H_n$ in (\ref{hatT}) are even, see Remark\;\ref{rem}) and
\begin{equation*}
R^{+}(\varepsilon,\varepsilon\nu,\beta)=R^{-}(\varepsilon,\varepsilon\nu,\beta)
\end{equation*}
by the evenness of $\bf Y$. In what follows, we will denote $R=R^{+}=R^{-}$ omitting the superindices. We also have the following
\begin{pro}\label{lv}
In the symmetric case (\ref{sym}) we have
\begin{equation*}
Q(\varepsilon,\varepsilon\nu,\beta)=R(\varepsilon,\varepsilon\nu,\beta).
\end{equation*}
\end{pro}
\no The proof of this proposition can be found in Appendix\;3.

\begin{remark}\label{complex}
The functions $Q$ and $R$ seem completely different, but in fact they are related because ${\bf g}^{(1)}$ and ${\bf g}^{(2)}$ are related, see (\ref{bfmgm1}).
\end{remark}
\no In the symmetric case $d_+=d_-$, $\gamma^2(0,\beta) \overline{d}_{+} d_{-}/\kappa=\Gamma/2$,
and formula (\ref{R}) for $\mathcal{R}$ implies that
\begin{equation}\label{SBW}
|\mathcal{R}|^2=\frac{\varepsilon^2\Gamma^2/4}{\delta^2+\varepsilon^2\Gamma^2/4}+O(\varepsilon)
\end{equation}
since $\delta W\Big(\ds\frac{\delta}{\varepsilon}\Big)= O(\varepsilon)$; a standard Breit-Wigner formula. This formula apparently means that $|\mathcal{R}|^2$ is symmetric with respect to the point $\delta=0$ and never vanishes (at least in the leading term).

\no Nevertheless, numerical data and theoretical investigations from, for example, \cite{Ship1, ChN} suggest that $|\mathcal{R}|^2$ vanishes at certain values of $\mu$ and the graph of $|\mathcal{R}|^2$ has a specific asymmetric form sometimes called ``the Fano resonance''.

\no In order to investigate these phenomena, we will calculate the zeros of $\mathcal{R}$ and  $\mathcal{T}$ following closely the presentation in \cite{Ship2}, although, as mentioned in the Introduction, our setting is rather different. To this end, we rewrite the formulas (\ref{mathcalR}) and (\ref{mathcalT}) in the following form: (cf.\cite{Ship1})
\begin{equation}\label{Sh}
\mathcal{R}=\frac{a}{\ell},\qquad\mathcal{T}=\frac{b}{\ell},
\end{equation}
where
\begin{equation}\label{call}
\ell=\nu-\gamma F,
\end{equation}
\begin{equation}\label{cala}
a=i k_0^{-1}\;\varepsilon\gamma\Big\lbrace (\nu-\gamma F)P^{+}+\gamma Q^2\Big\rbrace,
\end{equation}
\begin{equation}\label{calb}
b=\nu-\gamma F+i k_0^{-1}\;\varepsilon\gamma\Big\lbrace (\nu-\gamma F)P^{-}+\gamma Q^2\Big\rbrace,
\end{equation}
where the arguments $(\varepsilon\nu,\beta)$ and $(\varepsilon, \varepsilon\nu,\beta)$ are omitted in the functions $\gamma, k_0$ and $F, P^{\pm}, Q$, respectively.

\no Since the scattering is unitary, we have (see, e.g. \cite{Ship1})
\begin{equation}\label{Un}
|\ell|^2=|a|^2+|b|^2,\qquad \frac{a}{b}\in i\R.
\end{equation}
\begin{lemma}\label{ka1}
Let $\tilde{f}_{1}(\kappa)\neq0$ and $\tilde{f}_{0}(2\kappa)\neq0$, $\kappa=\sqrt{1-2\beta}$. Then the equations $a=0$ and $b=0$ possess, with respect to $\nu$, distinct real solutions $\nu_{a}=\nu_{a}(\varepsilon,\beta)$ and $\nu_{b}=\nu_{b}(\varepsilon,\beta)$ given by
\begin{equation}\label{Na}
\nu_a=\frac{(1-\beta)^2}{4\pi}\Bigg(\tilde{f}_0(0)-\Big(\tilde{f}_1(\kappa)\Big)^2\Big/\tilde{f}_0(2\kappa)\Bigg)+O(\varepsilon)
\end{equation}
\begin{equation}\label{Nb}
\nu_b=\frac{(1-\beta)^2}{4\pi}\tilde{f}_0(0)+O(\varepsilon).
\end{equation}
These solutions depend analytically on $\varepsilon$ and $\beta$ for $\beta\in(0,1/2)$ and small $\varepsilon$.
\end{lemma}
\no {\bf Proof.} We will prove the existence of $\nu_a$, for $\nu_b$ the proof is analogous. Obviously, $a=0$ when the expression $\mathcal{L}=(\nu-\gamma F)P^{+}+\gamma Q^2$ vanishes. This expression is analytic in $\nu, \varepsilon, \beta$ and by the Implicit Function Theorem possesses an analytic solution for $\nu$ since $\partial\mathcal{L}/\partial\nu\Big\vert_{\varepsilon=0}=P^{+}(0,0,\beta)=\tilde{f}_0(2\kappa)$.

\no Hence $\nu_a=\nu_a^{0}+\varepsilon \nu_a^{1}+\cdots$. Obviously, $\nu_a^{0}$ given by (\ref{Na}) is real. The equation for $\nu_a^{1}$ can be obtained by substituting $\nu=\nu_a^{0}+\varepsilon\nu_a^1$ in (\ref{cala}), dividing the resulting expression by $\varepsilon^2$ and passing to the limit as $\varepsilon\to 0$. Alternatively, the same equation can be obtained by considering the ratio $a/b$, dividing by $\varepsilon^2$ and passing to the same limit since
$$
b(\varepsilon,\nu,\beta)\Big\vert_{\nu=\nu_a^0+\varepsilon\nu_a^1}\xrightarrow[\varepsilon\to 0]{}-\frac{(1-\beta)^2}{4\pi}\Bigg(\Big(\tilde{f}_1(\kappa)\Big)^2\Big/\tilde{f}_0(2\kappa)\Bigg)\in\R,
$$
and the last expression does not vanish.
But this ratio is purely imaginary by (\ref{Un}) and hence the resulting equation for $\nu_a^1$ is purely real, and its solution is also real. By induction, all $\nu_a^{j}$ are real.

\begin{theorem}\label{va}
Let the conditions of Lemma \ref{ka1} be satisfied and let
\begin{equation}\label{Na phi}
\nu_a(0,\beta)>0,
\end{equation}
(see (\ref{omega2})); $\nu$ should be positive).
Then for $\nu=\nu_a$ we have $\mathcal{R}=0$ (total transmission) and for $\nu=\nu_b$ we have $\mathcal{T}=0$ (total reflection).
\end{theorem}
\no The proof follows immediately from Lemma \ref{ka1}.
\begin{remark}
Condition (\ref{Na phi}) is satisfied, for example, if $\tilde{f}_0(2\kappa)<0$ or if $\kappa$ is close to a zero of $\tilde{f}_0$. The latter case corresponds to the existence of an embedded trapped mode (see Section 7).
\end{remark}

\no The last theorem shows that indeed the graph of, for example, $|\mathcal{R}|^2$, cannot be symmetric with respect to the point $\delta=0$. In order to capture this asymmetry, we can preserve the term proportional to $\delta$ in (\ref{R}) in the derivation of the Breit-Wigner formula (\ref{SBW}).

\no It is easy to see from (\ref{R}) that with the same error bound $O(\varepsilon)$ we also have
\begin{equation}\label{F}
|\mathcal{R}|^2=\frac{\varepsilon^2(\delta q+\Gamma/2)^{2}}{\delta^2+\varepsilon^2\Gamma^{2}/4}+O(\varepsilon),
\end{equation}
where
$$
q=\frac{(1-\beta)^2}{4\pi}\;\frac{1}{\sqrt{1-2\beta}}\cdot\tilde{f}_0(2\kappa).
$$
This Fano type formula differs from (\ref{SBW}) by $O(\varepsilon)$ but captures the existence of a zero of $|\mathcal{R}|^2$ which lies outside the narrow (of order of $\varepsilon$) peak of the Breit-Wigner formula. The graphs of (\ref{SBW}) and (\ref{F}) are presented in Fig.\;2 for $\varepsilon=0.1$, $q=-1$ and $\Gamma=1$.
\begin{remark}\label{cre}
If $\tilde{f}_0(2\kappa)=0$, the reflection coefficient does not vanish in a neighborhood of size $\sim\varepsilon^{-1/2}$ of the point $\delta=0$ and hence we have just the Breit-Wigner resonance without the Fano ``dip'' (i.e., just the formula (\ref{SBW})). Nevertheless, the transmission coefficient does vanish  in this case at $\nu=\nu_b$.
\end{remark}

\vspace{1cm}
\begin{figure}[htbp]
\centering
\includegraphics[scale=0.3]{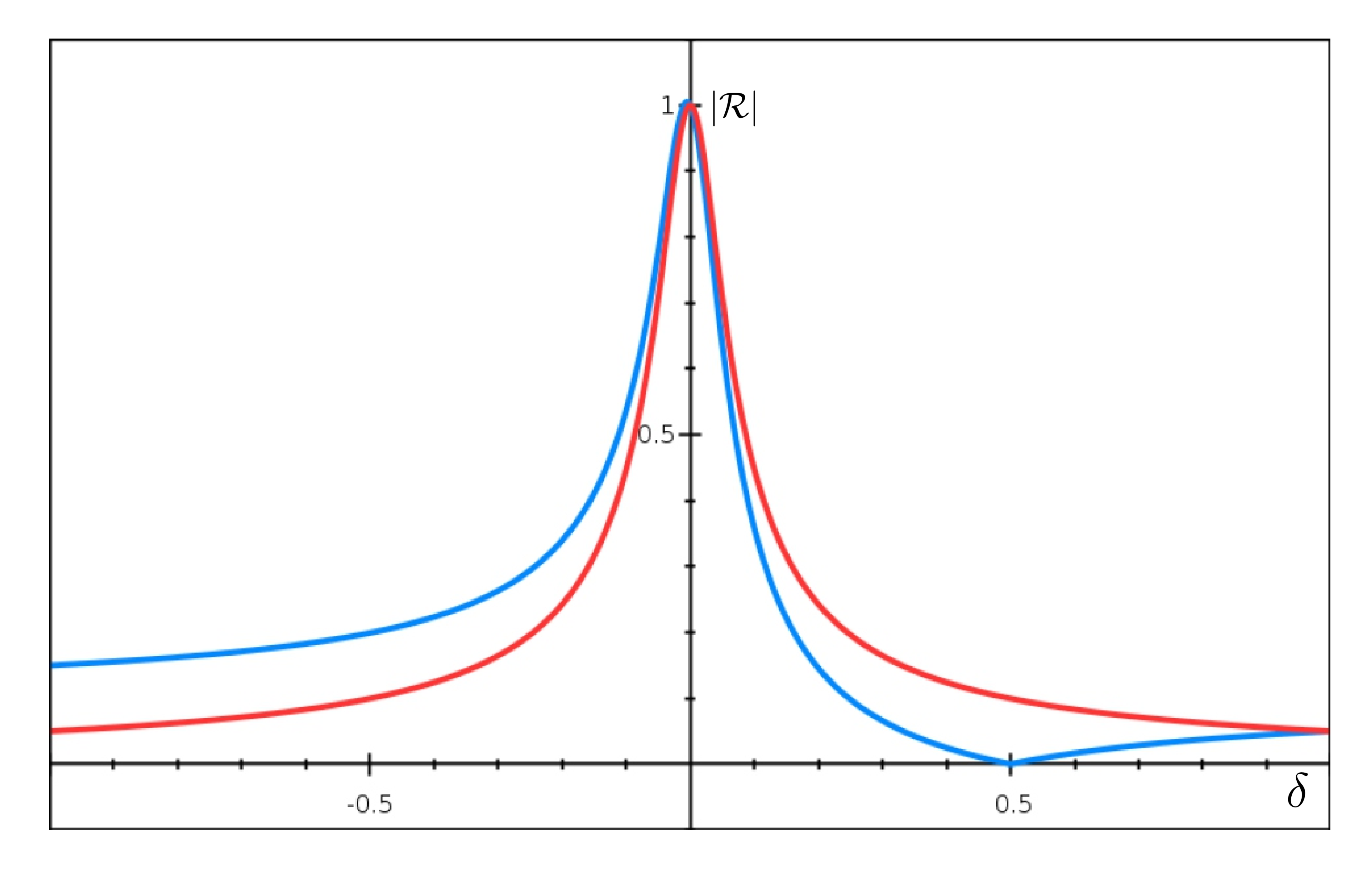}
\caption{Breit-Wigner (red) and Fano (blue) lines}
\end{figure}

\vspace{.5cm}

\section{Existence of embedded trapped mode and scattering.}
\setcounter{equation}{0}

In this section we will obtain the conditions for the existence of embedded RB waves. More exactly, we seek the values of the parameters $\beta$ (see, for example, (\ref{CP1})) and $\omega$ given by (\ref{omega2}) such that (\ref{hh2}), (\ref{bc1}) possesses a trapped mode. Note that in Sections 2-4 the parameter $\beta$ was an arbitrary fixed number such that $0<|\beta|<1/2$, and our considerations were valid for all these $\beta$. In this section we will find certain value of $\beta$ such that there exists a trapped RB wave. This value of $\beta$ corresponds to the fact that the root of the equation (\ref{eqm4}) is real. In this case the solution of the scattering problem is nonunique, as we will see in this section. We can restrict ourselves to the case of positive $\beta$ since if $\Psi$ is a trapped mode for $\beta$, then $\overline{\Psi}$ is a trapped mode for $-\beta$; thus here $0<\beta<1/2$.

\no Recall that we  consider the case of even $f$ satisfying (\ref{sym}).

\subsection{Embedded RB waves.}

Recall  that $\mu=\varepsilon\nu$. Dividing equation (\ref{eqm4}) by $\varepsilon$, we obtain
\begin{equation}\label{eqmu2}
\ell=\nu-\gamma(\varepsilon\nu,\beta) F(\varepsilon,\varepsilon\nu,\beta)=0,
\end{equation}
where $\gamma(\mu,\beta)$ and $F(\varepsilon,\mu,\beta)$ are analytic in $\varepsilon, \mu, \beta$ for small $\varepsilon, \mu$ and real $\beta$ such that $0<\beta<1/2$ (this follows directly from the explicit formulas (\ref{gamma}), (\ref{eqm5})). Generally speaking, the Taylor coefficients of the expansion of $F$ in powers of $\varepsilon, \mu$ are complex.

\no In this section we are interested in the case when these coefficients are real since then the solution $\nu(\varepsilon,\beta)=\mu_0(\varepsilon,\beta)/\varepsilon$ of (\ref{eqmu2}) ($\mu_0(\varepsilon,\beta))$ is defined in Section 4.1) is real and hence there exists a trapped mode. By definition (\ref{eqm5}), $F$ will have real Taylor coefficients when the expression
\begin{equation}\label{ex1}
\Big\langle\Big((1-\varepsilon\hat{T})^{-1}{\bf g}^{(2)}\Big)_{-1}\Big\rangle=\Big((1-\varepsilon\hat{T})^{-1}{\bf g}^{(2)},\bm{\ell}\Big)
\end{equation}
possesses this property. Here
\begin{equation}\label{6.3'}
({\bm \ell})_{n}:=\delta_{n}^{-1}
\end{equation}
where $\delta_n^m$ is the Kronecker symbol and $(\cdot, \cdot)$ denotes the scalar product
\begin{equation}\label{scp}
({\bf u}, {\bf v})=\sum_{-\infty}^{\infty}\int u_{n}(x) \overline{v}_{n}(x)\;dx,\quad {\bf u}, {\bf v}\in \mathcal{A}.
\end{equation}
Recall that
\begin{equation}\label{7.5}
{\bf Y}=(1-\varepsilon\hat{T})^{-1}{\bf g}^{(2)}~\Longleftrightarrow~(1-\varepsilon\hat{T}){\bf Y}={\bf g}^{(2)}.
\end{equation}
Obviously, if the Taylor coefficients of $ Y_{-1}$ are real, then the Taylor coefficients of (\ref{ex1}) are real.
Let
\begin{equation}\label{hT}
\hat{T}=\hat{T}_{1}+i\hat{T}_{2},
\end{equation}
where $\hat{T}_{1}$ and $\hat{T}_{2}$ are integral operator with real kernels when $\mu\in\R$ . If
\begin{equation}\label{6.6'}
\hat{T}_{2}{\bf Y}=0,
\end{equation}
then ${\bf Y}$ satisfies the equation
\begin{equation}\label{7.9'}
(1-\varepsilon\hat{T}_{1}){\bf Y}={\bf g}^{(2)},
\end{equation}
and hence has real coefficients, since ${\bf g}^{(2)}$
is real by (\ref{bfmgm1}) and (\ref{sym}), (\ref{i5}).
From (\ref{hT}), by the definition of $\hat{T}$ (\ref{hatT}), we obtain
\begin{equation}\label{7.6}
\Big(\hat{T}_{2}{\bf Y}\Big)_{m}=2\gamma\Big(\Im G_{0}\ast Y_{0}\Big) f_{m}
\end{equation}
(see properties of the Green functions $G_{m}$).
By (\ref{G_00}) the expression in the RHS of (\ref{7.6}) has the form
\begin{eqnarray*}
\begin{array}{ll}
\ds\frac{\gamma}{k_0}\Bigg(\ds\int\cos k_{0}(x-\xi)\;Y_{0}(\xi)\;d\xi\Bigg) f_{m}(x)=\ds\frac{\gamma}{k_0}\Bigg(\ds\int\cos k_0\xi\;Y_{0}(\xi)\;d\xi\Bigg) f_{m}(x)\cos k_{0}x,
\end{array}
\end{eqnarray*}
since $Y_0, G_0$ and $f_m$ are even. In fact, $f_m$ is even by (\ref{sym}) and (\ref{i5}). $G_0$ is even by (\ref{G_00}) and $Y_0$ is even by (\ref{7.5}) since $g^{(2)}$ is even by (\ref{bfmgm1}) and the operator $T$ preserves the evenness by (\ref{hatT}) and the evenness of all the kernels $H_{n}$, see Remark \ref{rem}. Hence, (\ref{7.6}) yields
\begin{equation}\label{ImT_2}
(\hat{T}_{2}{\bf Y})_{m}=\ds\frac{\gamma}{ k_0}\Bigg(\ds\int\cos \big(k_0\xi\big)\;Y_{0}(\xi)\;d\xi\Bigg) f_{m}(x)\cos k_{0}x.
\end{equation}
We see that (\ref{ex1}) will be real if the integral in (\ref{ImT_2}) is equal to zero. This condition can be rewritten in the form
\begin{equation}\label{02t}
R(\varepsilon,\varepsilon\nu,\beta)=0,
\end{equation}
where $R=R^{\pm}$ is defined in (\ref{PRpm}). Condition (\ref{02t}) can be considered as an equation for $\beta$ as a function of $\varepsilon, \varepsilon\nu$.
In the leading order with respect to $\varepsilon$ this equation reads
\begin{equation}\label{OC}
\tilde{f}_1\Big(\sqrt{1-2\beta}\Big)=0.
\end{equation}
We will assume that equation (\ref{OC}) possesses a real solution $\beta=\beta_{00}$ such that $0<\beta_{00}<1/2$.
\begin{remark}\label{Le_Pert}
There exist perturbations $f$ such that $\tilde{f}_1$ does not have real zeros, e.g. the Gaussian. Nevertheless we will see later that in some examples $\tilde{f}_1$ can have real zeros.
\end{remark}
\no In order to apply the Implicit Function Theorem to (\ref{02t}), we impose the condition
\begin{equation}\label{beta_0 var}
\frac{\partial}{\partial \beta} R(\varepsilon,\varepsilon\nu,\beta)\Big\vert_{(0,0,\beta_{00})}=\tilde{f}_{1}'\Big(\sqrt{1-2\beta_{00}}\Big)\neq0
\end{equation}
(in Example 6.11 below this condition is satisfied).

\medskip
\no Thus, if (\ref{OC}) and (\ref{beta_0 var}) hold, then (\ref{02t}) admits an analytic in $\varepsilon, \varepsilon\nu$ solution $\beta=\beta_0(\varepsilon,\varepsilon\nu)$. This solution has  real Taylor coefficients due to the fact that equation (\ref{02t}) is real, and hence defines (locally) a curve $\beta=\beta_0(\varepsilon,\varepsilon\nu)$ on the plane $(\beta,\nu)$ which in the leading order coincides  with the vertical line $\beta=\beta_{00}$, see Fig.\;3.

\vspace{1cm}
\begin{figure}[htbp]
\centering
\includegraphics[scale=0.3]{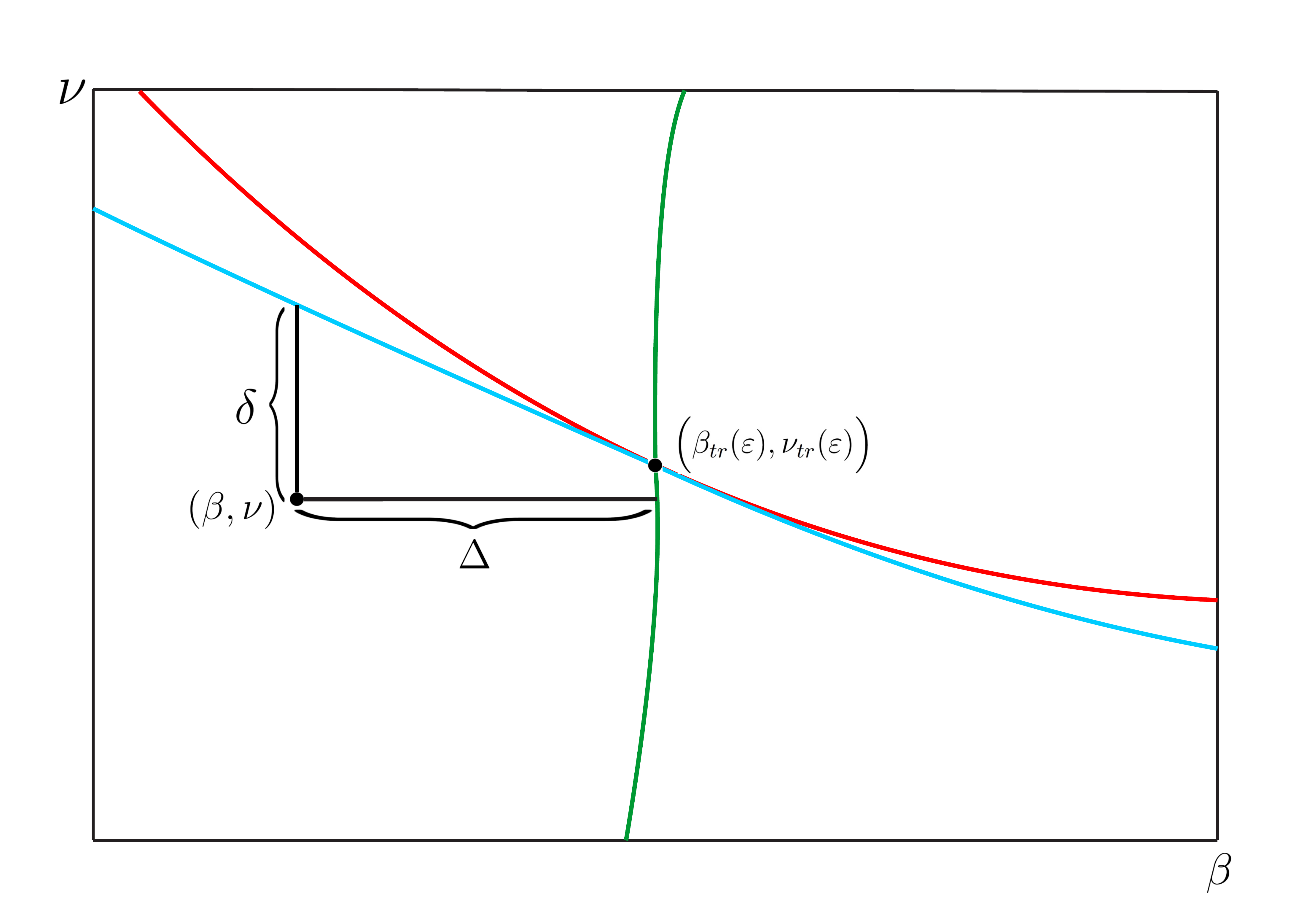}
\caption{The curves $\nu=\nu_a(\varepsilon,\beta)$ (red), $\nu=\Re\nu_0(\varepsilon, \beta)$ (blue) (this curve in the leading term coincides with $\nu=\nu_b(\varepsilon,\beta)$), and $\beta=\beta_0(\varepsilon,\varepsilon\nu)$ (green)}
\end{figure}

\vspace{.5cm}

\no Consider now equation (\ref{eqmu2}) on this curve. By the Implicit Function Theorem it possesses a root $\nu_{tr}(\varepsilon)$ on this curve, i.e.
\begin{equation}\label{r}
\nu_{t r}(\varepsilon)-\gamma\Big(\varepsilon\nu_{t r}(\varepsilon), \beta_0(\varepsilon,\varepsilon\nu_{t r}(\varepsilon))\Big) F\Big(\varepsilon,\varepsilon\nu_{t r}(\varepsilon), \beta_0(\varepsilon,\varepsilon\nu_{t r}(\varepsilon))\Big)=0.
\end{equation}
It is close to the point $\nu=\nu_0(0,\beta_{00})$, and, since (\ref{eqmu2}) is real on $\beta=\beta_0$, this root is real. Denote
\begin{equation}\label{beta_t}
\beta_{tr}(\varepsilon)=\beta_0\Big(\varepsilon,\varepsilon\nu_{t r}(\varepsilon)\Big).
\end{equation}
Obviously
\begin{equation}\label{r_t}
\nu_{tr}(\varepsilon)-\gamma\Big(\varepsilon\nu_{t r}(\varepsilon), \beta_{t r}(\varepsilon)\Big) F\Big(\varepsilon, \varepsilon\nu_{t r}(\varepsilon), \beta_{t r}(\varepsilon)\Big)=0\qquad\beta_{00}=\beta_{tr}(0).
\end{equation}
\begin{remark}
$\nu_{tr}(\varepsilon)$ and $\beta_{tr}(\varepsilon)$ can be interpreted also as the solution of the system
$$
\nu-\gamma(\varepsilon\nu,\beta) F(\varepsilon,\varepsilon\nu,\beta)=0,\quad Q(\varepsilon,\varepsilon\nu,\beta)=0.
$$
\end{remark}
\no At the point $\Big(\beta_{tr}(\varepsilon), \nu_{tr}(\varepsilon)\Big)$ there exists an embedded trapped mode; we prove it in the following Theorem\;\ref{kappa(beta)} below. To formulate it, we will need some notation.

\no Define $\Psi(x,y,\mu)$ by formula (\ref{Psi}) replacing $\beta$ by $\beta_{tr}(\varepsilon)$, $\mu$ by $\varepsilon\nu_{tr}(\varepsilon)$ and defining $\Psi_m$ as
\begin{equation}\label{Psi_0=G_0ast A_0}
\Psi_{m}=G_{m}\ast A_{m}, \quad m\in\Z,
\end{equation}
with
\begin{equation}\label{bf A}
{\bf A}={\bf Y}=(1-\varepsilon\hat{T})^{-1} {\bf g}^{(2)}
\end{equation}
(see (\ref{7.5})). Note the difference between (\ref{Psi_0=G_0ast A_0}) and (\ref{Psi_0,Psi_m}) for $m=0$.
Denote the obtained function as
\begin{equation}\label{psi(math{B})}
\Psi(x,y,\varepsilon).
\end{equation}
Now we formulate the main theorem of this section.
\begin{theorem}\label{kappa(beta)}
Let $\kappa_0=\kappa(\beta_{00})=\sqrt{1-2\beta_{00}}\in(0,1)$ be such that
\begin{equation*}
\tilde{f}_{1}(\kappa_0)=0, \quad \tilde{f}'_{1}(\kappa_0)\neq0.
\end{equation*}
Then the function $\Psi(x,y,\varepsilon)$ defined by (\ref{psi(math{B})}) is a finite energy solution of problem (\ref{hh2}), (\ref{bc1}) with $\omega^2=\omega_1^2-\varepsilon^2\nu_{tr}^2(\varepsilon)$, and $\beta=\beta_{tr}(\varepsilon)$.
\end{theorem}
\begin{remark}\label{matcal}
The leading terms of the Taylor expansions for $\nu_{tr},~ \beta_{tr}$ imply formulas (\ref{mf})  from the Introduction.
\end{remark}
{\bf Proof of  Theorem\;\ref{kappa(beta)}.}
{\bf I.} The fact that $\Psi(x,y,\varepsilon)$ satisfies (\ref{hh2}), (\ref{bc1}) follows directly from the construction of the functions ${\bf A}, \Psi$, (see (\ref{Psi}), (\ref{Psi_0=G_0ast A_0}), (\ref{bf A}). Indeed, substituting (\ref{Psi_0=G_0ast A_0}) in (\ref{Psi}) and then in (\ref{hh2}) we see that in order that $\Psi$ be a solution to (\ref{hh2}) ${\bf A}$ should satisfy
$$
(1-\varepsilon\hat{T}){\bf A}=\frac{\omega^2}{4\pi\nu_{tr}(\varepsilon)} \langle A_{-1}\rangle{\bf g}^{(2)}.
$$
This equality is identically satisfied iff $\ds\frac{\omega^2}{4\pi\nu_{tr}(\varepsilon)} \langle A_{-1}\rangle=1$ by (\ref{bf A}). The last equality is true since $\nu_{tr}$ solves (\ref{eqmu2}) and
\begin{equation*}
\langle A_{-1}\rangle=\Big\langle\Big((1-\varepsilon\hat{T})^{-1} {\bf g}^{(2)}\Big)_{-1} \Big\rangle
\end{equation*}
by (\ref{bf A}).

\no {\bf II.} Let us show that $\Psi(x,y,\varepsilon)$ decays as $|x|\to\infty$. Obviously, it is sufficient to prove this for $\Psi_{0}(x,y,\varepsilon)$, since $\Psi_{m}, m\neq0$, possess this property by (\ref{Psi_0,Psi_m}), (\ref{G_m}) and the fact that $\supp A_{n}$ is compact by (\ref{eA}). For $\Psi_0$ we have
$$
\Psi_0=G_0\ast A_0=G_0\ast Y_0
$$
by (\ref{Psi_0=G_0ast A_0}). Hence,
\begin{equation*}
  \Psi_0(x)=\frac{i}{2k_0}\int e^{ik_0|x-\xi|}\;Y_0(\xi)\;d\xi=\left\{
 \begin{array}{ll}
  \ds\frac{i}{2k_0} e^{ik_0x} R\big(\varepsilon,\varepsilon\nu_{tr}(\varepsilon),\beta_{tr}(\varepsilon)\big), & x\gg 1,\\\\
\ds\frac{i}{2k_0} e^{-ik_0x} R\big(\varepsilon,\varepsilon\nu_{tr}(\varepsilon),\beta_{tr}(\varepsilon)\big), & x\ll -1,
 \end{array}
 \right.
 \end{equation*}
and hence, by (\ref{02t}), $\Psi_0(x)=0$~ for~ $|x|\gg1$.~~~\bo

\subsection{Scattering in a neighborhood of embedded RB wave.}

Now we will consider the scattering problem in the case
\begin{equation*}
\beta=\beta_{t r}(\varepsilon),\quad \mu=\varepsilon\nu_{t r}(\varepsilon),
\end{equation*}
where $\beta_{t r}(\varepsilon)$ and $\nu_{t r}(\varepsilon)$ are defined by (\ref{r_t}).
The scattering problem (\ref{Psi1}) does not have a unique solution when there exists an embedded trapped mode since a  multiple of this mode can always be added to the solution of (\ref{Psi1}). In this subsection we will show that nevertheless the reflection and transmission coefficients are well-defined. We begin with the proof of the fact that
if the orthogonality condition (\ref{02t}) holds then the RHS of (\ref{eqA_1}) vanishes and hence the scattering problem is soluble albeit nonuniquely.
Indeed, in this case formulas (\ref{eA}), (\ref{bfg_2}) for the solution of the scattering problem still hold with $C$ an arbitrary constant. The last statement follows from Proposition\;\ref{lv}  which means that the right-hand side of (\ref{eqA_1}) vanishes and the fact that $\mu=\varepsilon\nu_{t r}(\varepsilon)$, that is, the second factor in the left-hand side of (\ref{eqA_1}) also vanishes by (\ref{eqmu2}).

\no Thus, we come to  the following
\begin{theorem}\label{X}
Let $\beta=\beta_0(\varepsilon, \varepsilon\nu)$. Then

\no {\bf (1)} if $\nu=\nu_{t r}(\varepsilon)$, then the scattering problem (\ref{Psi1}) has a nonunique solution given by (\ref{eA}) and (\ref{bfg_2}) with an arbitrary constant $C$ and the transmission and reflection coefficients are uniquely defined by (\ref{R,T}).

\no {\bf (2)} if $\nu\neq\nu_{t r}(\varepsilon)$, then the scattering problem (\ref{Psi1}) has a unique solution given by (\ref{eA}) and (\ref{bfg_2}) with $C=0$.

\no In both cases $\mathcal{R}=O(\varepsilon)$, $\mathcal{T}=1+O(\varepsilon)$ (i.e., neither Breit-Wigner nor Fano resonances are present on the curve $\beta=\beta_0(\varepsilon,\varepsilon\nu)$, see Fig.\;3).
\end{theorem}
\no {\bf Proof.} Consider the function
\begin{equation*}
{\bf A}:=2\varepsilon\gamma(1-\varepsilon\hat{T})^{-1}\;{\bf h},
\end{equation*}
where
\begin{equation*}
{\bf h}={\bf g}^{(1)}+\frac{C}{2\mu}\;{\bf g}^{(2)}
\end{equation*}
and $C$ is an arbitrary constant.
Let us prove that under the condition (\ref{eqmu2}) $\langle A_{-1}\rangle=C$. Indeed,
\begin{equation*}
\langle A_{-1}\rangle=2\varepsilon\gamma\Bigg\lbrace \Big\langle \Big((1-\varepsilon \hat{T})^{-1} {\bf g}^{(1)}\Big)_{-1}\Big\rangle+\frac{C}{2\mu} \Big\langle \Big((1-\varepsilon \hat{T})^{-1} {\bf g}^{(2)}\Big)_{-1}\Big\rangle\Bigg\rbrace.
\end{equation*}
The first summand in the curly brackets is equal to $0$ by Proposition\;\ref{lv} and the second is equal to $\ds\frac{2\pi C}{\varepsilon\omega^2}$ by (\ref{eqmu2}). In fact by (\ref{eqm5}) $\Bigg\langle \Bigg(\Big(1-\varepsilon T\Big)^{-1} {\bf g}^{(2)}\Bigg)_{-1}\Bigg\rangle=F\Big(\varepsilon, \varepsilon\nu_{t r}(\varepsilon), \beta_{t r}(\varepsilon)\Big)$ which is equal to $\nu_{t r}(\varepsilon)/\gamma(\varepsilon\nu_{tr},\beta_{tr})$ by (\ref{r_t}).
Hence ${\bf h}={\bf g}$ and ${\bf A}$ satisfies (\ref{eA}), therefore $\Psi$ given by (\ref{Psi}) and (\ref{Psi_0,Psi_m}) satisfies the Helmholtz equation.
The last assertion of the theorem follows from  formulas (\ref{R,T}) since $C=0$ in (\ref{eqC}).
\begin{remark}\label{raw}
When the orthogonality condition is not satisfied (that is, $\beta\neq\beta_0$), Theorem\;\ref{teo} states that $\mathcal{R}$ can become large. In contrast, in the case under consideration our last result shows that the reflection coefficient is small.
\end{remark}

 Introduce new coordinates $\delta,\Delta$ with the origin at $(\beta_{t r},\nu_{t r})$ by the formulas
\begin{equation}\label{delta,Delta}
\delta=\nu-\Re\nu_0(\varepsilon,\beta),\qquad \Delta=\beta-\beta_0(\varepsilon,\varepsilon\nu).
\end{equation}
This change of the variables is nondegerate since its Jacobean is equal to $1+O(\varepsilon)$.

Consider the curves $\nu=\nu_a$ and $\nu=\nu_b$ defined in Lemma\;\ref{ka1}. Clearly, they pass through the point $(\beta_{t r},\nu_{t r})$ since $a=b=0$ at this point (see formulas (\ref{cala}), (\ref{calb})).

\no From (\ref{cala}), (\ref{calb}) we obtain that these curves intersect tangentially at the point $(\beta_{tr}, \nu_{tr})$ and are transversal to the curve $\beta=\beta_0(\varepsilon,\varepsilon\nu)$. In coordinates $(\delta,\Delta)$ this easily follows from (\ref{ell}), (\ref{Im nu_0}) and (\ref{Q2}) below.

 We will be interested in the asymptotics of $\mathcal{R}$ and $\mathcal{T}$ in a neighborhood of     $(\beta_{t r},\nu_{t r})$.
Consider the asymptotics of $\mathcal{R}$, $\mathcal{T}$ for small values of $\delta,\Delta$.

\no Let
\begin{equation*}
\alpha:=\frac{2\gamma_0}{\kappa_0}|\tilde{f}'_{1}(\kappa_0)|^2,\qquad q:=\frac{\gamma_0}{\kappa_0}\tilde{f}_0(2\kappa_0),\qquad \gamma_0:=\gamma(0,\beta_{00}).
\end{equation*}
\begin{theorem}
$\mathcal{R}$ and $\mathcal{T}$ admit the asymptotics
\begin{eqnarray}\label{mHR}
\mathcal{R}&=&\frac{i\varepsilon\Big(\delta q+\alpha\Delta^2+(\delta+\Delta^2)O(\varepsilon+|\Delta|)\Big)\Big(1+O(\varepsilon+|\Delta|)\Big)}{\delta-i\varepsilon\alpha\Delta^2}
                            ,    \\  \nonumber\\ \label{mHT}
\mathcal{T}&=&\frac{\Big(\delta +(\delta+\Delta^2)O(\varepsilon+|\Delta|)\Big)\Big(1+O(\varepsilon+|\Delta|)\Big)}{\delta-i\varepsilon\alpha\Delta^2}.
                                   \end{eqnarray}
\end{theorem}
\no {\bf Proof.}
Let us calculate the asymptotics of $\ell$ in (\ref{call}). We have
\begin{equation}\label{ell}
\ell=\nu-\gamma(\varepsilon,\beta) F(\varepsilon, \varepsilon\nu,\beta)=\Big(1+O(\varepsilon)\Big)\Big(\nu-\nu_0(\varepsilon,\beta)\Big)=\Big(1+O(\varepsilon)\Big)\Big(\delta-i\Im\nu_0(\varepsilon,\beta)\Big).
\end{equation}
Since $\Im\nu_0(\varepsilon,\beta)$ is positive (see \cite[Theorem\;15]{Ship1}) and vanishes for $\beta=\beta_0(\varepsilon,\varepsilon\nu)$ we have
\begin{equation}\label{Im nu_0}
\Im\nu_0(\varepsilon,\beta)=\varepsilon\Delta^2\Big(\alpha+O(\varepsilon+|\Delta|)\Big).
\end{equation}
Here $O(\varepsilon+|\Delta|)$ is uniform with respect to $\delta$, since $\delta$ enters $\Im\nu_0$  through the product $\varepsilon\delta$ (see (\ref{delta,Delta})). Further, we have by (\ref{ell}) and (\ref{Im nu_0}), and using an analog of (\ref{den}),
\begin{equation}\label{1/ell}
\frac{1}{\ell}=\frac{1+O(\varepsilon)}{\nu-\nu_0}=\frac{1+O(\varepsilon+|\Delta|)}{\delta-i\varepsilon\alpha\Delta^2}
\end{equation}
uniformly in $\delta$. Indeed,
\begin{equation*}
\begin{array}{lll}
\ds\frac{1}{\delta-i\varepsilon\alpha\Delta^2}-\ds\frac{1}{\delta-i\varepsilon\alpha\Delta^2+\varepsilon\Delta^2 O(\varepsilon+|\Delta|)}=\ds\frac{\varepsilon\Delta^2 O(\varepsilon+|\Delta|)}{\Big(\delta-i\varepsilon\alpha\Delta^2+\varepsilon\Delta^2 O(\varepsilon+|\Delta|)\Big)\Big(\delta-i\varepsilon\alpha\Delta^2\Big)}\\\\
\qquad \qquad\qquad=\ds\frac{1}{\delta-i\varepsilon\alpha\Delta^2}\cdot\ds\frac{O(\varepsilon+|\Delta|)}{\ds\frac{\delta}{\varepsilon\Delta^2}-i\alpha+O(\varepsilon+|\Delta|)}
\end{array}
\end{equation*}
and
$$
\ds\frac{1}{\ds\frac{\delta}{\varepsilon\Delta^2}-i\alpha+O(\varepsilon+|\Delta|)}=O(1)
$$
since $\alpha\neq0$ by (\ref{beta_0 var}).
Further, by (\ref{cala})
\begin{equation}\label{a}
a=ik_0^{-1}\varepsilon\gamma\Big\lbrace (\nu-\gamma F) P^{+}+\gamma Q^2\Big\rbrace.
\end{equation}
Since $Q\Big(\varepsilon, \varepsilon\nu, \beta_0(\varepsilon, \varepsilon\nu)\Big)=0$, we have
\begin{equation}\label{Q2}
\begin{array}{lll}
\gamma(\varepsilon\nu,\beta)\;Q^2(\varepsilon,\varepsilon\nu,\beta)=\gamma\Big(\varepsilon (\delta+\Re\nu_0),\Delta+\beta_0\Big)\;Q^2\Big(\varepsilon, \varepsilon(\delta+\Re\nu_0),\Delta+\beta_0\Big)\\\\
\qquad\qquad\qquad= \Delta^2\Bigg(\gamma_0\Big(\tilde{f}'_{1}(\kappa_0)\Big)^2+O(\varepsilon+|\Delta|)\Bigg),
\end{array}
\end{equation}
$$P^{+}(\varepsilon,\varepsilon\nu,\beta)=\tilde{f}_0(2\kappa_0)+O(\varepsilon+|\Delta|).$$
Substituting in (\ref{a}) and using (\ref{ell}), (\ref{Im nu_0}), we obtain
\begin{equation*}
\begin{array}{lll}
a&=&\ds\frac{i\varepsilon\gamma_0}{\kappa_0}\Bigg(\Big(\delta-i\varepsilon\Delta^2\big(\alpha+O(\varepsilon+|\Delta|)\big)\Big)\Big(\tilde{f}_0(2\kappa_0)+O(\varepsilon+|\Delta|)\Big)+\\\\\
&+& \Delta^2\Big(\gamma_0\big(\tilde{f}'_{1}(\kappa_0)\big)^2+O(\varepsilon+|\Delta|)\Big)\Bigg)\Big(1+O(\varepsilon+|\Delta|)\Big)\\\\
&=& \ds\frac{i\varepsilon\gamma_0}{\kappa_0}\Bigg[\delta\tilde{f}_0(2\kappa_0)+\Delta^2\gamma_0\Big(\tilde{f}'_{1}(\kappa_0)\Big)^2+\Big(\delta+\Delta^2\Big) O(\varepsilon+|\Delta|)\Bigg]\Big(1+O(\varepsilon+|\Delta|)\Big).
\end{array}
\end{equation*}
Hence for $\mathcal{R}$ we obtain (\ref{mHR}) using (\ref{1/ell}). Similarly, we obtain (\ref{mHT}). ~~$\bo$
\medskip

\no Clearly, (\ref{mHR}) implies that
\begin{equation}\label{add R}
\mathcal{R}=\frac{i\varepsilon\alpha\Delta^2}{\delta-i\varepsilon\alpha\Delta^2}+O(\varepsilon+|\Delta|),
\end{equation}
but of course  formula (\ref{mHR}), being a multiplicative asymptotics, takes into account the zeros of $\mathcal{R}$, while (\ref{add R}), the standard Breit-Wigner formula, being additive,  does not.
\begin{remark}
If
\begin{equation}\label{f02k}
\tilde{f}_0(2\kappa_0)=0,
\end{equation}
 we can guarantee that $\mathcal{R}$ does not vanish, for sufficiently small $\varepsilon$, in a $\Delta$-neighborhood of the point $\delta=0$ (note that if $\tilde{f}_0(2\kappa_0)\neq0$, then $\mathcal{R}=0$ at a point whose distance to $\delta=0$ is of order of $\Delta^2$). As in Remark \ref{cre}, the transmission coefficient vanishes at $\nu=\nu_b$.
\end{remark}
\no We conclude with two examples which illustrate conditions (\ref{OC}) and (\ref{f02k}).
\begin{example}
(rectangular barrier). Let
\begin{equation}\label{fxy}
f(x,y)=g(x)(1+\cos y)
\end{equation}
where $g(x)$ is the characteristic function of the interval $|x|<a$. We have
\begin{equation*}
\tilde{f}_0(\xi)=\frac{4\pi}{\xi}\sin a\xi,\quad \tilde{f}_{1}(\xi)=\tilde{f}_0(\xi)/2.
\end{equation*}
Condition (\ref{OC}) reads as $\kappa=\pi n/a,~n=1,2,\cdots$. Obviously, for sufficiently large $a$, we have $0<\kappa<1$ and hence the corresponding $\beta\in(0,1/2)$. Simultaneously, $\tilde{f}_0(2\kappa)=0$, so we will have the purely Breit-Wigner resonance in a neighborhood of this value of $\beta$ ($\mathcal{R}$ does not vanish in this neighborhood). On the other hand, if $\kappa\neq\pi n/a$, then $\tilde{f}_0(2\kappa)\neq0$; we do not have a trapped mode but we have the Fano resonance ($\mathcal{R}=0$ at the point $\nu=\nu_a$).
\end{example}

\begin{example}
(parabolic barrier). Let $g(x)=1-x^2/a^2$ for $|x|<a$ and $g(x)=0$ elsewhere. We have
$$
\tilde{f}_0(\xi)=\frac{8\pi}{a^2\xi^3}(\sin a\xi-a\xi\cos a\xi),\quad \tilde{f}_1(\xi)=\tilde{f}_0(\xi)/2.
$$
Condition (\ref{OC}) reads as $\tan a\xi=a\xi$. Thus (\ref{OC}) is satisfied if $\kappa={x}/{a}$ where $x$ is the root of the equation $\tan x=x$ in the interval $\Big({\pi}/{2},{3\pi}/{2}\Big)$. For sufficiently large $a$, we have $0<\kappa<1$. It is easy to see that $\tilde{f}_0(2\kappa)\neq0$; thus, we have the Fano resonance in a neighborhood of the corresponding value of $\beta$ ($\mathcal{R}$ vanishes at $\nu=\nu_a$).
\end{example}
\begin{remark}
In spite of the fact that function $f$ is not smooth and hence does not satisfy our conditions of smoothness (\ref{C1}), this is not essential since one can consider a smoothed perturbation obtained by means of a convolution of $g$ with a suitable kernel instead of $g$.
\end{remark}

\section{Appendix 1. Proof of Lemma \ref{lImvar_0neq0}. }
\setcounter{equation}{0}

 For the vectors ${\bf H}$, ${\bf A}$, ${\bf f}$ with components
\begin{equation}\label{H,A,f}
H_{m}=H_{m}(x),~~A_{m}=A_{m}(x)~~{\rm and}~~ f_{m}=f_{m}(x)
\end{equation}
denote
\begin{equation}\label{ast,circledast}
\Big({\bf H}\ast{\bf A}\Big)_{m}:=H_{m}\ast A_{m};~~~\Big({\bf f}\circledast {\bf A}\Big)_{m}=\sum_{n\in\Z} f_{m-n} A_{n}.
\end{equation}
{\bf Proof of Lemma \ref{lImvar_0neq0}}. Since $\mu_{0}(\varepsilon)$ satisfies (\ref{eqm4}), $\mu_{0}(\varepsilon)$ admits, by the Implicit Function Theorem, the  expansion (\ref{mu0-0}). Hence,
\begin{equation}\label{mu3}
\mu-\varepsilon\gamma(\mu,\beta)\Big\langle \Big(g^{(2)}+\varepsilon Tg^{(2)}\Big)_{-1}\Big\rangle+O(\varepsilon^{3})=0.
\end{equation}
Since
$
\omega^{2}=\omega^{2}_1-\mu^2,~~~\omega^{2}_1=(1-\beta)^{2},
$
substituting (\ref{mu0-0}) into (\ref{mu3}), we obtain, equating to zero the coefficients of powers of $\varepsilon$ and $\varepsilon^2$,
\begin{equation}\label{a1}
a_{1}-\gamma(0,\beta)\; \Big\langle \Big({\bf g}^{(2)}\Big)_{-1}\Big\rangle=0, \quad a_{2}-\gamma(0,\beta)\; \Big\langle \Big(\hat{T}\big\vert_{\mu=0}\;{\bf g}^{(2)}\Big)_{-1}\Big\rangle=0.
\end{equation}
Formula (\ref{a1}) implies that $a_1$ is real.
Let us calculate $a_2$.
We have
\begin{eqnarray*}
\begin{array}{ll}
\Big(\hat{T}\big\vert_{\mu=0}\;{\bf A}\Big)_{m}=2\gamma(0,\beta) \Big({\bf f}\circledast {\bf H}\ast{\bf A}\Big)_{m}=2\gamma(0,\beta)\ds\sum_{n\in\Z} f_{m-n}\;H_{n}\ast A_{n}.
\end{array}
\end{eqnarray*}
Since $\Big({\bf g}^{(2)}\Big)_{m}=f_{m+1}$ by (\ref{bfmgm1}),
\begin{eqnarray}\label{Tg2}
\begin{array}{ll}
\Big\langle\Big(\left.\hat{T}\right|_{\mu=0}{\bf g}^{(2)}\Big)_{-1}\Big\rangle=2\gamma(0,\beta)\Big\langle\ds\sum_{n\in\Z} f_{-n-1}\;\big(H_{n}\ast f_{n+1}\big) \Big\rangle\\\\\hspace{2.4truecm}=2\gamma(0,\beta)\ds\sum_{n\in\Z}\ds\int\int f_{-n-1}(\xi)\;H_n(\xi-\eta) f_{n+1}(\eta)\;d\xi d\eta \\\\\hspace{2.4truecm}=2\gamma(0,\beta)\ds\sum_{n\in\Z}\ds\int\int H_{n}(\xi-\eta) f_{n+1}(\eta)\;\overline{f}_{n+1}(\xi)\;d\xi d\eta
\end{array}
\end{eqnarray}
by (\ref{i5}).
Denoting the integral in the RHS of (\ref{Tg2}) by $Z_{n}$, we have
$$
\overline{Z}_{n}=2\gamma(0,\beta)\ds\sum_{n\in\Z}\ds\int\int H_{n}(\xi-\eta)\;\overline{f}_{n+1}(\xi)\;f_{n+1}(\eta)\;d\xi d\eta=Z_{n}, \quad n\neq0,
$$
since $H_{n}$ is real and even by (\ref{Hn}) $\forall n\neq0$, and
for $n=0$,
$
\Im H_{0}(x)=(\cos k_{0}x)/2k_{0}
$
by (\ref{Gr}), (\ref{Hn}). Hence we obtain
\begin{equation*}
\Im a_2=\gamma(0,\beta)\;\Im\Big\langle\Big(\hat{T}\big\vert_{\mu=0}\;{\bf g}^{(2)}\Big)_{-1}\Big\rangle=\ds\frac{\gamma^2(0,\beta)}{\kappa}\ds\int\int \cos \kappa(\xi-\eta) f_{1}(\xi)\;\overline{f}_{1}(\eta)\;d\xi d\eta.
\end{equation*}
Further,
\begin{eqnarray*}
\begin{array}{ll}
 \ds\int\cos \kappa(\xi-\eta) f_{1}(\xi)\;\overline{f}_{1}(\eta)\;d\xi d\eta=\ds\frac{1}{2}\ds\int e^{i\kappa(\xi-x)}f_{1}(\xi)\;\overline{f}_{1}(\eta)\; d\xi d\eta+\ds\frac{1}{2}\ds\int e^{-i\kappa(\xi-\eta)}f_{1}(\xi)\;\overline{f}_{1}(\eta)\; d\xi d\eta\\\\\qquad\qquad\qquad\hspace{.1truecm}=
 \ds\frac{1}{2}\Big(\tilde{f}_1(-\kappa)\;\tilde{\overline{f}}_1
 (\kappa)+\tilde{f}_1(\kappa)\;\tilde{\overline{f}}_1(-\kappa)\Big)=\ds\frac{1}{2}\Big(|\tilde{f}_1(-\kappa)|^2+|\tilde{f}_1(\kappa)|^2\Big)
\end{array}
\end{eqnarray*}
since $\tilde{\overline{f}}_1(\kappa)=\overline{\tilde{f}_1}(-\kappa)$.
The lemma is proven.  ~~$\bo$

\section{Appendix\;2. Proof of formula (\ref{Lag}).}
\setcounter{equation}{0}

Denote $K(\varepsilon,\mu,\beta)=\gamma(\mu,\beta) F(\varepsilon,\mu,\beta)$. We have
$$
\nu-K(\varepsilon, \varepsilon\nu,\beta)=\nu-\nu_0(\varepsilon,\beta)-\Big(K(\varepsilon,\varepsilon\nu,\beta)-K(\varepsilon,\varepsilon\nu_0,\beta)\Big)
$$
since $\nu_0$ solves $\nu-K(\varepsilon,\varepsilon\nu,\beta)=0$, see (\ref{eqnu}). By the Lagrange formula
$$
K(\varepsilon, \varepsilon\nu,\beta)-K(\varepsilon, \varepsilon\nu_0,\beta)=\varepsilon(\nu-\nu_0)\int\limits_{0}^1 K_{\mu}\Big(\varepsilon, \varepsilon\theta(\nu-\nu_0)+\varepsilon\nu_0,\beta\Big)\;d\theta.
$$
Since $K_{\mu}$ is bounded, we obtain (\ref{Lag}) since $\nu-\nu_0(\varepsilon,\beta)=\delta-i\varepsilon\Gamma/2+O(\varepsilon^2)$ by Lemma \ref{lImvar_0neq0} and since $\varepsilon|\delta|=O(\sqrt{\varepsilon})$ by (\ref{delta}).

\section{Appendix\;3. Proof of Proposition \ref{lv}.}
\setcounter{equation}{0}

We will need the following technical proposition.
\begin{lemma}\label{la2}
Let $\mathcal{H}$ be a space with a scalar product $(\cdot,\cdot)$ and $\hat{Q}_1$, $\hat{Q}_2$ linear  operators $\mathcal{H}\to\mathcal{H}$ (in general noncommuting), $\hat{Q}_{1}^{*}=\hat{Q}_{1}$. Then $\forall u, v\in\mathcal{H},~n\in\N_{0}$
\begin{equation*}
\Big((\hat{Q}_1 \hat{Q}_2)^{n}\;(\hat{Q}_1 u), v\Big)=\Big((\hat{Q}_2 \hat{Q}_1)^{n} u,\hat{Q}_1 v\Big)=\Big(u, (\hat{Q}_1 \hat{Q}^{*}_2)^{n} (\hat{Q}_1 v)\Big),
\end{equation*}
where $\hat{Q}^{*}_2$ is the adjoint operator.
\end{lemma}
\no {\bf Proof.} The proof of this lemma is evident.~~~$\bo$

\no {\bf Proof of Proposition \ref{lv}}. For ${\bf A}\in\mathcal{A}$ denote
\begin{equation}\label{6.22'}
\hat{Q}_1{\bf A}={\bf f}\circledast{\bf A}, \quad\hat{Q}_{2}{\bf A}={\bf H}\ast{\bf A},
\end{equation}
where ${\bf H}, {\bf A}, {\bf f}$ are defined by (\ref{H,A,f}), see (\ref{ast,circledast}).
Then the operator $\hat{T}$ defined by (\ref{hatT}) is
\begin{equation}\label{hat{T}}
\hat{T}=2\gamma\;\hat{Q}_1 \hat{Q}_2.
\end{equation}
Note that by (\ref{sym}) $\hat{Q}_1$ is self-adjoint as easily follows from the definition (\ref{ast,circledast}).

\no Denote the shift operator
\begin{equation*}
(\hat{S}{\bf f})_{m}=f_{m+1}.
\end{equation*}
Let
\begin{equation}\label{6.22''}
\hat{T}:=\hat{T}_1+i\hat{T}_2,
\end{equation}
where $\hat T_{1,2}$ are defined by (\ref{hatT}) with $H_n$ substituted by $\Re H_n$ or $\Im H_n$, respectively.
Using Lemma\;\ref{la2} and (\ref{6.22'}) we have (the scalar product is defined in (\ref{scp}))
\begin{eqnarray}\label{6.23'}
\begin{array}{ll}
\Big(\hat{T}^{n}{\bf g}^{(1)},{\bm \ell}\Big)=\Big(\hat{T}^{n}({\bf f}\circledast(e^{ik_0x}{\bf m})),\hat{S} {\bf m}\Big)=\Big((2\gamma\;\hat{Q}_{1}\hat{Q}_{2})^{n}\Big({\bf f}\circledast(e^{ik_0x}{\bf m})\Big), \hat{S}{\bf m}\Big)=\\\\\hspace{2.1truecm}=\Big((2\gamma\;\hat{Q}_{2}\hat{Q}_{1})^{n}(e^{ik_0x}{\bf m}),{\bf f}\circledast \hat{S}{\bf m}\Big)=\Big(e^{ik_0 x}{\bf m}, (\gamma\;\hat{Q}_{1}\hat{Q}^{*}_{2})^{n}({\bf f}\circledast \hat{S}{\bf m})\Big).
\end{array}
\end{eqnarray}
Let us prove that
\begin{equation}\label{gammaQ_1}
2\gamma\;\hat{Q}_1 \hat{Q}^{*}_2=\hat{T}_1-i\hat{T}_2.
\end{equation}
We have by (\ref{hat{T}}) that
$$
\hat{T}=\hat{T}_1+i\hat{T}_2=2\gamma \hat{Q}_1 \hat{Q}_2= 2\gamma \hat{Q}_1\Big(\hat{Q}_2^{(1)}+i\hat{Q}_2^{(2)}\Big).
$$
Hence,
$$
2\gamma\;\hat{Q}_1 \hat{Q}^{*}_2=2\gamma\;\hat{Q}_1\Big(\hat{Q}_2^{(1)}-i\hat{Q}_2^{(2)}\Big)=2\gamma\;\hat{Q}_1 \hat{Q}_2^{(1)}-2i\gamma\;\hat{Q}_1 \hat{Q}_2^{(2)}=\hat{T}_1-i\hat{T}_2,
$$
since $\hat{Q}^{*}_2=\hat{Q}_2^{(1)}-i\hat{Q}_2^{(2)}$ and $\hat{Q}_2^{(1)}$ and $\hat{Q}_2^{(2)}$ are self-adjoint operators. So, (\ref{gammaQ_1}) holds. Thus, by (\ref{6.23'})
$$
\Big(\hat{T}^{n}\;{\bf g}^{(1)},{\bm\ell}\Big)=\Big(e^{ik_0 x}{\bf m}, (\hat{T}_1-i\hat{T}_2)^{n}(f\circledast \hat{S}{\bf m})\Big),\quad n\in\N.
$$
Hence,
\begin{equation}\label{ast ast}
\Big((1-\varepsilon(\hat{T}_1+i\hat T_2))^{-1}\;{\bf g}^{(1)},{\bm\ell}\Big)=\Big(e^{ik_0 x}{\bf m}, (1-\varepsilon(\hat{T_1}-i\hat T_2))^{-1}\;{\bf g}^{(2)}\Big).
\end{equation}
In fact
$$
\sum_{n}\Big(\hat{T}^{n}\;{\bf g}^{(1)},{\bm\ell}\Big)\varepsilon^{n}=\sum_{n}\Big(e^{ik_0 x}{\bf m}, (\hat{T}_1-i\hat{T}_2)^{n}(f\circledast \hat{S}{\bf m})\Big)\varepsilon^{n}.
$$
So (\ref{ast ast}) holds. Since ${\bf f}\circledast \hat{S}{\bf m}={\bf g}^{(2)}$ by (\ref{bfmgm1})
\begin{equation}\label{ast}
\Big((1-\varepsilon\hat{T})^{-1}\;{\bf g}^{(1)},{\bm\ell}\Big)=\Bigg(e^{ik_0 x}{\bf m},\Big(1-\varepsilon(\hat{T}_1-i\hat{T}_2)\Big)^{-1}\;{\bf g}^{(2)}\Big)\Bigg).
\end{equation}
Note that ${\bf Y}$ defined by (\ref{7.5}) satisfies
$$
\overline{{\bf Y}}=\Big(1-\varepsilon(\hat{T}_1-i\hat{T}_2)\Big)^{-1} {\bf g}^{(2)}
$$
and hence
$$
\Big((1-\varepsilon\hat{T})^{-1}\;{\bf g}^{(1)}, {\bm\ell}\Big)=\Big(e^{ik_0 x} {\bf m},\overline{{\bf Y}}\Big)=\Big({\bf Y}, e^{ik_0 x} {\bf m}\Big)= R(\varepsilon, \varepsilon\nu,\beta)
$$
since ${\bf Y}$ is even. Proposition\;\ref{lv} is proven. ~~$\bo$

\section{Conclusion.}

We have constructed explicit formulas for the solution of scattering problem for the two-dimensional Helmholtz equation with a weak periodic perturbation of the wave speed. These formulas have the form of  infinite convergent series in powers of the small parameter characterizing the perturbation and are valid for almost all values of the ``wavenumber'' when a certain orthogonality condition is not satisfied. In particular, we obtained formulas of Breit-Wigner type for the reflection and transmission coefficients; these formulas imply an anomalous behavior of these coefficients for some frequencies.

 When the orthogonality condition is satisfied, the solution of the scattering problem is nonunique and the problem possesses an embedded Rayleigh-Bloch mode whose frequency is explicitly calculated.

Moreover, we indicated the existence, under certain conditions, of anomalies in the transmission and reflection coefficients of the Fano type (asymmentry with respect to frequency) and also proved the existence of frequencies for which the system possesses total transmission and/or reflection.

Finally, by means of a fortunate choice of coordinates, we  obtained the asymptotics of the reflection and transmission coefficients uniformly with respect to the distance to the trapped mode.

\end{document}